\documentclass[letterpaper]{JHEP3}

\usepackage{amsmath,amssymb,amsfonts}
\usepackage{bm,bbm}
\usepackage{epsfig}

\newcommand{\beq}{\begin{eqnarray}}
\newcommand{\eeq}{\end{eqnarray}}
\newcommand{\nn}{\nonumber}

\newcommand{\eql}[1]{\label{eq:#1}}
\newcommand{\eq}[1]{(\ref{eq:#1})}
\newcommand{\figl}[1]{\label{fig:#1}}
\newcommand{\fig}[1]{\ref{fig:#1}}
\newcommand{\Fig}[1]{Figure \ref{fig:#1}}
\newcommand{\Secl}[1]{\label{sec:#1}}
\newcommand{\Sec}[1]{Section\ \ref{sec:#1}}
\newcommand{\Appl}[1]{\label{app:#1}}
\newcommand{\App}[1]{Appendix \ref{app:#1}}
\newcommand{\bitem}{\begin{itemize}}
\newcommand{\eitem}{\end{itemize}}

\newcommand{\al}{\alpha}
\newcommand{\be}{\beta}
\newcommand{\ga}{\gamma}

\newcommand{\de}{\delta}
\newcommand{\ep}{\epsilon}
\newcommand{\la}{\lambda}
\newcommand{\sg}{\sigma}
\newcommand{\tht}{\theta}

\newcommand{\cL}{\mathcal{L}}
\newcommand{\cO}{\mathcal{O}}
\newcommand{\cC}{\mathcal{C}}

\newcommand{\lt}{\left}
\newcommand{\rt}{\right}
\newcommand{\wba}[1]{{\overline{#1}}}
\newcommand{\sla}[1]{{\raise.15ex\hbox{$/$}\kern-.57em #1}}
\newcommand{\Sla}[1]{\kern0.12em{\raise.15ex\hbox{$/$}\kern-.74em #1}}
\newcommand{\del}{\partial}

\newcommand{\fr}[2]{\frac{#1}{#2}}

\newcommand{\gsim}{\gtrsim}
\newcommand{\lsim}{\lesssim}

\newcommand{\fb}{\text{fb}}
\newcommand{\Gev}{\text{GeV}}

\newcommand{\SU}{{\mathop{\rm SU}}}
\newcommand{\U}{{\mathop{\rm {}U}}}
\newcommand{\SUC}{\SU(3)_\text{C}}
\newcommand{\SUL}{\SU(2)_\text{L}}
\newcommand{\UY}{\U(1)_\text{Y}}

\newcommand{\hpi}{{\tilde{\pi}}}
\newcommand{\hrho}{{\tilde{\rho}}}
\newcommand{\hpiT}{\hpi_\text{T}}
\newcommand{\hpiD}{\hpi_\text{D}}
\newcommand{\hpiS}{\hpi_\text{S}}

\title{The LHC Phenomenology of Vectorlike Confinement}

\author{Can Kilic\\
Department of Physics \& Astronomy, Rutgers University, Piscataway, NJ 08854, USA\\
Email: \email{kilic@physics.rutgers.edu}}
\author{Takemichi Okui\\
Department of Physics, Florida State University, Tallahassee, FL 32306, USA\\
Email: \email{okui@hep.fsu.edu}}

\abstract{We investigate in detail the LHC phenomenology of ``vectorlike confinement'', where the Standard Model is augmented by a new confining gauge
interaction and new light fermions that carry vectorlike charges under both the Standard Model and the new gauge group. If the new interaction confines at
the TeV scale, this framework gives rise to a wide range of exotic collider signatures such as the production of a vector resonance that decays to a pair
of collider-stable charged massive particles (a ``di-CHAMP'' resonance), to a pair of collider-stable massive colored particles (a ``di-R-hadron''
resonance), to multiple photons, $W$s and $Z$s via two intermediate scalars, and/or to multi-jet final states. To study these signals at the LHC, we set
up two benchmark models: one for the di-CHAMP and multi-photon signals, and the other for the di-R-hadron and multi-jet signals. For the
di-CHAMP/multi-photon model, Standard Model backgrounds are negligible, and we show that a full reconstruction of the spectrum is possible, providing
powerful evidence for vectorlike confinement. For the di-R-hadron/multi-jet model, we point out that in addition to the di-R-hadron signal, the rate of
the production of four R-hadrons can also be sizable at the LHC. This, together with the multi-jet signals studied in earlier work, makes it possible to
single out vectorlike confinement as the underlying dynamics.}


\preprint{RUNHETC-2010-01}

\begin{document}

%%%%%%%%%%%%%%%%%%%%%%
\section{Introduction}
\Secl{intro}
%%%%%%%%%%%%%%%%%%%%%%
Our exploration of physics beyond the Standard Model (SM) is about to enter an extremely exciting era with the startup of the LHC. We know that there must
exist new physics beyond the SM to explain unmistakable observational evidence such as neutrino oscillations, a variety of astrophysical effects pointing
to dark matter, the cosmological baryon asymmetry necessitating CP violation beyond the SM, and the nearly scale-invariant primordial density fluctuations
indicative of inflation in the early universe. Despite decades of effort by theorists, however, there is no convergence on a single theory that is even
remotely as compelling {\em and} specific as the SM was at the time when it guided the discoveries of the $W$ and $Z$ bosons at the SPS and the top quark
at the Tevatron \cite{MTT}. In the absence of such convergence, a productive approach for the LHC is to keep an open mind to which field-theoretical
structures may exist beyond the SM, and broadly explore their experimental manifestation at the LHC.

In collider phenomenology, the two most interesting mechanisms for the production of a new particle are resonant production and pair production. While the
collider signatures of virtually every proposed extension of the SM follow from these two primary processes (and subsequent decays), it is by no means
true that all possibilities have been explored exhaustively, even though the existence of a new particle and its couplings to the SM are severely
constrained by precision electroweak and flavor measurements. In this context, the scenario of ``vectorlike confinement'' proposed in Ref.\ \cite{VC} is
particularly interesting, as it
\begin{itemize}
{\item gives rise to both of these two primary processes from a simple fundamental lagrangian,}
{\item evades the most restrictive constraints automatically such as direct dilepton and dijet resonance searches as well as precision
flavor and electroweak measurements,}
{\item and introduces a new phenomenological class of decay modes of the produced particles down to SM final states.}
\end{itemize}
Let us touch upon each of these properties briefly: vectorlike confinement augments the SM at the TeV scale in the manner that QCD augments QED at the GeV
scale. A new confining gauge interaction (``hypercolor'') and new vectorlike fermions (``hyperquarks'') are added to the SM. Hyperquarks are assumed to be
light compared to the hypercolor confinement scale, analogously to the $u$, $d$ and $s$ quarks, which are light compared to the QCD confinement scale. The
hyperquarks, once pair produced, rapidly form bound states due to hypercolor confinement, allowing for both resonant and pair productions of ``hypermesons''
at the LHC, analogous to the resonant and pair productions of the $\rho$ mesons and pions at a sub-GeV $e^+$-$e^-$ collider.

While there are many possible models of vectorlike confinement, the existences of the spin-1 bound state (analogous to the QC $\rho$ meson) and
a pseudoscalar bound state (analogous to the QCD pion) are completely general.  Like $\rho \to \pi\pi$ in QCD, the spin-1 resonances dominantly decay
into a pair of pseudoscalar bound states.  These features are explained at length in Ref.\ \cite{VC} and will be reviewed below.

The fact that the hyperquarks appear in vectorlike representations of the SM gauge groups makes it trivial to evade precision electroweak constraints,
while the fact that the SM gauge interactions are the dominant connection to the new physics ensures a minimal impact on any flavor observable. Finally,
the fact that the spin-1 resonances predominantly decay into pseudoscalar pairs severely weakens the constraints coming from resonance searches
in dilepton or dijet channels.

The pseudoscalars carry SM charges and some (but not necessarily all) of them decay to a pair of SM gauge bosons. Thus, when a pair of such pseudoscalars
are produced either from Drell-Yan processes or from the decay of a spin-1 resonance, the final state contains four SM gauge bosons. When the four gauge
bosons are electroweak gauge bosons ($W$s, $Z$s and $\ga$s), the event can be quite spectacular, and such events can be utilized to reconstruct both the
parent spin-1 resonance and the two intermediate pseudoscalar resonances.
\footnote{A special model in which the four gauge bosons are always gluons was studied in Ref.\ \cite{Coloron-1} and shown to have a significant discovery
potential in existing ($\approx 1\>\fb^{-1}$) Tevatron 4-jet data. Ref.\ \cite{Coloron-2} extended this analysis to the context of the LHC, demonstrating
the discovery potential in the multi-jet channel.}
A second phenomenological feature that is quite generic in vectorlike confinement models is the existence of charged and/or colored massive pseudoscalars
that are stable on collider time scales. If colored, such a particle will hadronize under QCD, thereby forming a massive stable hadron (an ``R-hadron''),
which will carry a net electric charge an $\cO(1)$ fraction of time. For the charged, color-neutral long-lived pseudoscalars, we will adopt the
commonly-used name of ``CHAMPs'' (CHArged Massive Particles). The CHAMPs/R-hadrons in vectorlike confinement can be pair-produced from Drell-Yan processes
as well as decays of the new spin-1 resonances, giving rise to the unusual experimental feature of a resonance in CHAMP pairs (a ``di-CHAMP resonance'')
or in R-hadron pairs (a ``di-R-hadron resonance'').

While similar in terms of the field contents, vectorlike confinement utilizes an entirely different hierarchy of energy scales compared to
``quirk models'' \cite{Quirks}, where the new fermions are assumed to be much heavier than the confinement scale of the new gauge interaction. This
difference in energy scales is essential and the phenomenology of vectorlike confinement is very much different from that of the quirk scenario. The
phenomenology of vectorlike confinement also differs from that of the ``hidden valley'' scenario \cite{HV} where the new fermions are ``hidden'', i.e.,
completely neutral under all SM interactions; in contrast, the hyperquarks in vectorlike confinement carry SM charges. The common property that vectorlike
confinement shares with the quirk and hidden valley scenarios (as well as QCD) is that a very simple microscopic lagrangian gives rise to a surprisingly
rich phenomenology exhibited by a variety of bound states with a wide range of lifetimes.

The purpose of this paper is complementary to that of Ref.\ \cite{VC}; instead of having a qualitative survey of generic features of vectorlike
confinement models, we aim to set up specific benchmark models suitable for numerical simulations, and study the CHAMP and R-hadron final states as well
as those of four SM gauge bosons produced from the decay of intermediate pseudoscalars. These benchmark models are intended to help develop an intuition
for the signatures of vectorlike confinement. We will introduce two benchmark models; the signatures of the first one include CHAMPs as well as
multi-photons (we will focus our attention on a $3\ga + W$ final state). We will study a few points in the parameter space of this model to show that the
reconstruction of the parent spin-1 resonance from CHAMP pairs as well as the multi photons is quite promising at the LHC, constituting a powerful probe
of vectorlike confinement. The second benchmark has R-hadrons instead of CHAMPs. Here, the di-R-hadron resonance signal has a much larger cross-section
compared to the di-CHAMP signal. This benchmark also has four gauge boson signals, but most of the time these consist of four gluons, obscured by large SM
backgrounds. The LHC discovery of the pseudoscalar resonances in this channel was shown to be possible in Ref.\ \cite{Coloron-2}, although the parent
spin-1 resonance is more difficult to reconstruct from multi-jets. In this benchmark we will focus on the R-hadrons instead, as they are much easier to
discover. In particular, we will point out a novel and spectacular signature of the production of {\em four} R-hadrons coming from the decays of a pair of
spin-1 resonances.

This paper is organized as follows. In \Sec{recap}, we review the framework and generic phenomenological features of vectorlike confinement. In
\Sec{benchmarks}, we study the phenomenology of the two benchmark models in detail and analyze various observables of the signals. In \Sec{conc}, we
conclude. We also include an appendix where we summarize the full theoretical details of the benchmark models.

%%%%%%%%%%%%%%%%%%%%%%%%%%%%%%%%%%%%%%%%
\section{A Vectorlike Confinement Recap}
\Secl{recap}
%%%%%%%%%%%%%%%%%%%%%%%%%%%%%%%%%%%%%%%%
The fundamental Lagrangian of a vectorlike confinement model has a very simple form:
\beq
  \cL = \cL_\text{SM} - \frac{1}{4} H_{\mu\nu} H_{\mu\nu}
       +\sum_i \lt( \wba{\psi}_i \,i\sla{\del} \psi_i
                   -m_i\, \wba{\psi}_i \psi_i
                   -g_\text{HC}\, \wba{\psi}_i \,\Sla{H} \psi_i
                   -g_\text{SM}\, \wba{\psi}_i \,\Sla{A}_\text{SM} \psi_i
               \rt)  ,
\eql{VC}
\eeq
where $H_{\mu\nu} H_{\mu\nu}$ is the kinetic term for a new strong confining gauge interaction, {\em hypercolor}, described by a gauge field $H$, while
$A_\text{SM}$ collectively denotes SM gauge fields. Reflecting our interest in the LHC phenomenology, we choose the scale of hypercolor confinement to be
$\cO(1)$ TeV. $\psi$s are new fermions, {\em hyperquarks}, and different $\psi$s can have different SM gauge quantum numbers and masses. Hyperquarks are
assumed to interact via both hypercolor and SM gauge interactions, but note that there are no direct couplings of hyperquarks to SM fermions or the Higgs
boson.
\footnote{Gauge invariance allows no direct {\em renormalizable} couplings between hyperquarks and SM fermions.  In general we expect small
nonrenormalizable couplings, e.g.\ $\psi_i\psi_j f_a f_b$ with SM fermions $f_a$ and $f_b$, but the absence of excessive flavor/CP violations beyond the
SM implies that such couplings must be very small, in fact so small that they have no effects on collider phenomenology, as discussed in detail in Ref.\
\cite{VC}. We will therefore ignore nonrenormalizable couplings in this paper.}
Furthermore, SM gauge interactions of hyperquarks are {\em vectorlike} (i.e.\ no $\ga_5$ appearing in \eq{VC}).

The lagrangian \eq{VC} has the same structure as the QED-QCD lagrangian (i.e.\ the SM at GeV$\lsim E \ll M_\text{W}$):
\beq
  \cL_\text{QED-QCD}
 = \cL_\text{QED} - \frac{1}{4} G_{\mu\nu} G_{\mu\nu}
       +\sum_i \lt( \bar{q}_i \,i\sla{\del} q_i
                   -m_i\, \bar{q}_i q_i
                   -g_\text{QCD}\, \bar{q}_i \,\sla{G} q_i
                   -g_\text{QED}\, \bar{q}_i \,\Sla{A} q_i
               \rt)  ,
\eql{QED-QCD}
\eeq
where $G$ and $q_i$ denote the gluon and quarks, while $A$ denotes the photon and $\cL_\text{QED}$ is the QED lagrangian for the leptons and photon.  The
analogy is quite accurate --- quarks interact via both color and QED interactions; their couplings to the photon is vectorlike; and there are no direct
renormalizable coupling between quarks and leptons. Using this analogy, we can qualitatively understand all the important phenomenological features of
vectorlike confinement. In the di-CHAMP benchmark of \Sec{dichamp}, the analogy is actually useful even quantitatively.

Many lagrangians similar to \eq{VC} have been extensively studied in the literature. For example, in the long tradition of $\U(1)'$(or $Z'$)
phenomenology, $H$ is a new massive $\U(1)$ gauge boson, with or without new vectorlike fermions $\psi$.  Here, the phenomenology drastically differs from
vectorlike confinement since $\U(1)$ does not lead to confinement. Hidden valley models \cite{HV} and quirk models \cite{Quirks} are especially
well-studied scenarios with a new confining gauge interaction and vectorlike fermions. In hidden valley models, all $\psi$s are ``hidden'', i.e., they
have no SM gauge interactions unlike the $\psi$s in vectorlike confinement. The $\psi$s in quirk scenarios can feel SM gauge interactions, but the $\psi$
masses are assumed to be much larger than the scale of hypercolor confinement, while they are assumed to be smaller in vectorlike confinement. (See below
for more on this assumption.)

%%%%%%%%%%%%%%%%%%%%%%%%%%%%%%%%%%%%%%%%%%%%%%%%
\subsection{Vector and Pseudo-scalar Resonances}
%%%%%%%%%%%%%%%%%%%%%%%%%%%%%%%%%%%%%%%%%%%%%%%%
In the QED-QCD analogy, a $\rho$ meson can be resonantly produced from an $e^+$-$e^-$ collision, where the $e^+$-$e^-$-$\rho$ coupling originates from the
$\rho$'s mixing with a photon.  Similarly, in vectorlike confinement, a massive spin-1 {\em hyper-rho meson}, $\hrho$, can be resonantly produced at
hadron colliders from a $q$-$\bar{q}$ collision, where the $q$-$\bar{q}$-$\hrho$ coupling arises from the $\hrho$'s mixing with a SM gauge boson.
The $\hrho$ mass will be near the hypercolor confinement scale $\sim \cO(1)$ TeV.
Once produced, the $\hrho$ promptly decays to $\hpi\hpi$ where $\hpi$s are light pseudoscalar {\em hyperpions}, analogously to $\rho \to \pi\pi$ in QED-QCD.
(See \Fig{rho-pi-pi}.) Alternatively, $\hpi$ pairs can be produced directly through ``Drell-Yan'' processes if they carry SM quantum numbers. (See
\Fig{Drell-Yan}.) Decays of $\hpi$s constitute a rich phenomenology and deserve separate subsections (see Sections \ref{sec:pi-long} and
\ref{sec:pi-short}); below we list other important characteristic properties of $\hrho$s and $\hpi$s relevant for collider phenomenology:%
\footnote{For a derivation and a more detailed discussion of these results, we refer the reader to Ref.\ \cite{VC}.}
\DOUBLEFIGURE[t]{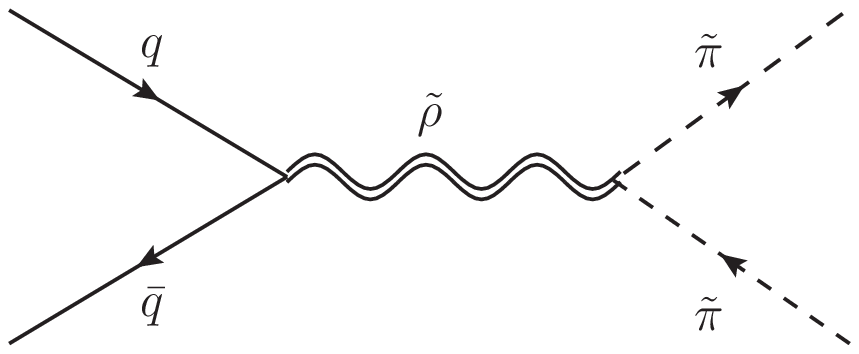, width=1.0\linewidth} {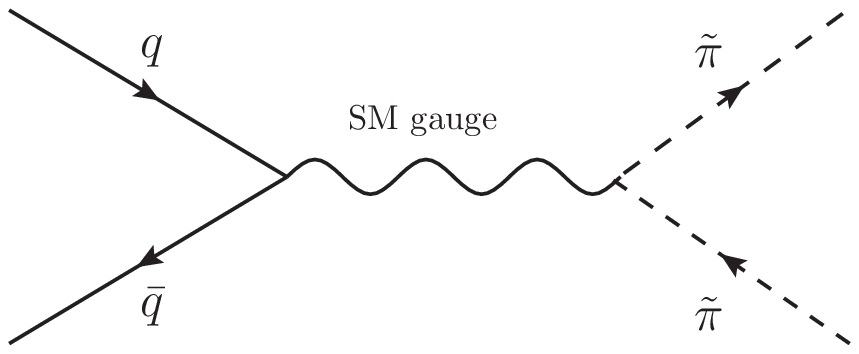, width=1.0\linewidth} {\figl{rho-pi-pi}The $\hpi$ production via a $\hrho$
resonance.} {\figl{Drell-Yan}The Drell-Yan $\hpi$ production.}
\bitem
\item[(i)]{There can be different kinds of $\hrho$s, e.g.\ the one that mixes with a gluon, the one that
mixes with a $W^+$, etc. Hence, we will often use notations such as $g'$, $W'^{+}$, etc.\ to distinguish among this variety, while using $\hrho$ to refer
to them collectively.  Similarly, there are many kinds of $\hpi$s with different SM quantum numbers.}
\item[(ii)]{Couplings of $\hrho$s to SM fermions are flavor-blind, i.e.\ they do not distinguish generations.
In particular, the $g'$ in vectorlike confinement differs from the Kaluza-Klein gluon in certain Randall-Sundrum models where it has stronger couplings to
the 3rd generation fermions \cite{KKgluons}.}
\item[(iii)]{Couplings of $\hpi$s to SM fermions are highly suppressed,
hence playing no role in the production of $\hpi$s.  This is because unlike $\hrho$s, which have spin one, $\hpi$s have spin zero and thus
cannot mix with SM bosons to acquire a sizable coupling to SM fermions.}
\item[(iv)]{We regard the dominance of the $\hrho \to \hpi\hpi$ mode
as part of the {\em definition} of vectorlike confinement. This is equivalent to assuming that hyperquark masses are sufficiently small to keep the phase
space open for $\hrho \to \hpi\hpi$.  This in particular means that the decays of $\hrho$s to SM fermions are subdominant, liberating $\hrho$ from
constraints from resonance searches in dijet and di-lepton channels. The dominance of $\hrho \to \hpi\hpi$ is therefore a distinguishing feature of
vectorlike confinement from well-studied $Z'$, $W'$, and $g'$, which predominantly decay to SM fermions.}
\item[(v)]{Since hyperquark masses are much smaller than the $\hrho$ masses,
and SM gauge loop corrections to the $\hrho$ masses are also small, all $\hrho$s are practically degenerate in mass.  This is analogous to the near
degeneracy of $\rho^0$ and $\rho^\pm$ in QED-QCD.  Hence, we use a single common value $m_\hrho$ for the $\hrho$ masses. In contrast, $\hpi$s exhibit a
rich spectrum reflecting their SM quantum numbers and the underlying hyperquark masses.} \eitem
The resonant $\hrho$ production followed by $\hrho \to \hpi\hpi$ is {\em the} signature process of vectorlike confinement.  Depending on how the $\hpi$s
decay, this process subsequently manifests itself in a range of final states, which we will now discuss.

%%%%%%%%%%%%%%%%%%%%%%%%%%%%%%%%%
\subsection{CHAMPs and R-hadrons}
\Secl{pi-long}
%%%%%%%%%%%%%%%%%%%%%%%%%%%%%%%%%
The lagrangian \eq{VC} has a conserved ``$\psi_i$ number'' for each $i$. Therefore, hyperpions consisting of two different hyperquarks carry a nonzero
$\psi$ number and cannot decay to SM particles, since SM particles do not carry $\psi$ numbers.  Therefore, the lightest $\hpi$ with given $\psi$ numbers
is stable and will be referred as a ``$\hpi_\text{long}$''. Note, however, that the conservation of $\psi$ numbers is just an accidental property of the
renormalizable interactions in \eq{VC}; nonrenormalizable interactions can violate $\psi$ numbers, thereby letting $\hpi_\text{long}$s decay. (See
\App{fundamental} for explicit forms of the $\hpi$-decaying nonrenormalizable operators for the benchmark models introduced below.) Interestingly, as
discussed in Ref.\ \cite{VC}, without introducing extra flavor symmetry in the framework, the absence of excessive flavor/CP violations beyond the SM
requires such nonrenormalizable operators to be so small that $\hpi_\text{long}$s can be considered stable on the collider time scale. A QED-QCD analogue
of $\hpi_\text{long}$s is the charged pion $\pi^\pm$, which is long-lived ($c\tau \sim 10$ m) due to its nonzero ``$u$ number'' (or ``$d$ number''), which
is accidentally conserved in the renormalizable lagrangian \eq{QED-QCD} but is violated by nonrenormalizable 4-fermion interactions arising from
integrating out the $W$ boson.

Note that there will exist a $\hpi_\text{long}$ whenever there is more than one $\psi$ in the theory. In this sense, the existence of $\hpi_\text{long}$s
is a very generic feature of the vectorlike confinement framework. Moreover, since $\hpi_\text{long}$s are made of two different hyperquarks, they
generically have nontrivial SM quantum numbers. A $\hpi_\text{long}$ that carries electric charge will appear as a CHAMP, behaving like a massive ``muon''
in a detector. A colored $\hpi_\text{long}$, on the other hand, will hadronize with SM quarks/gluons to form a massive long-lived hadron, i.e.\ an
R-hadron.

The fact that CHAMPs/R-hadrons can be produced from the decay of a $\hrho$ resonance (\Fig{rho-pi-pi}) in addition to the ordinary Drell-Yan channel
(\Fig{Drell-Yan}) is an intrinsic feature of vectorlike confinement, with the immediate consequence that the invariant mass distribution of those
CHAMP/R-hadron pairs exhibits a resonant peak at $m_\hrho$. Furthermore, in hadron colliders, the pair production of $\hpi_\text{long}$s, through
Drell-Yan processes as well as $\hrho$ decays, proceeds through a spin-1 $s$-channel intermediate state (except in the case of colored $\hpi$s produced
from a $g$-$g$ initial state), which shifts the production away from the threshold region. Therefore the $\hpi_\text{long}$s tend to be more boosted than
CHAMPs/R-hadrons in other extensions of the SM, such as the long-lived scalar taus in some gauge-mediated supersymmetric models. Ref.\ \cite{Todd-Jay}
proposed a very efficient technique to differentiate boosted CHAMPs from muons at the LHC, with an increased sensitivity and a promising discovery
potential even in the early LHC running.

%%%%%%%%%%%%%%%%%%%%%%%%%%%%%%%%%%%%%%%%%
\subsection{Multiple Gauge Boson Signals}
\Secl{pi-short}
%%%%%%%%%%%%%%%%%%%%%%%%%%%%%%%%%%%%%%%%%
A hyperpion consisting of a $\psi_i$ and a $\wba{\psi}_i$ with the same $i$ carries no net $\psi$ numbers and decays promptly to a pair of SM gauge
bosons; we will refer to such hyperpions as $\hpi_\text{short}$s.  $\hpi_\text{short}$s are qualitatively analogous to the $\pi^0$ in QED-QCD, which
carries neither $u$ number nor $d$ number and decays promptly ($c\tau \sim 10$ nm) to $\ga\ga$. Similarly, a $\hpi_\text{short}$ decays quickly to a pair
of SM gauge bosons. Unlike $\hpi_\text{long}$s, the existence of a $\hpi_\text{short}$ does not require the theory to contain more than one $\psi$,
so it is a guaranteed feature of vectorlike confinement. For example, Refs.\ \cite{Coloron-1, Coloron-2} specialize the case where there is
only one $\psi$, which carries color but no electroweak charges, thereby giving rise to a $\hpi_\text{short}$ that decays to $gg$.

While $\hpi$s are spin-0 particles carrying SM charges like sfermions in supersymmetric models, $\hpi_\text{short}$s decay to SM gauge boson pairs, quite
unlike the sfermions. Even in supersymmetric models with broken $R$ parity \cite{brokenR},
sfermions decay to SM fermions rather than to gauge bosons. One may think that
$\hpi_\text{short}$s that decay to $WW$ and $ZZ$ are similar to a Higgs boson. However, $\hpi_\text{short}$s have a branching fraction to $\ga\ga$ that is
comparable to those to $WW$ and $ZZ$! Therefore, the difference is quite dramatic --- in vectorlike confinement scenarios with an electroweak-charged
$\psi$, the LHC will observe multiple photon events with a cross-section consistent with electroweak {\em pair} production. This also means that
irreducible SM backgrounds are very small.  If multiple-photon events are observed at the LHC, the reconstruction of the intermediate $\hpi$ resonances as
well as the primary $\hrho$ resonance can provide strong evidence for vectorlike confinement, and we will lay out an experimental strategy for that below.

Finally, a $\hpi_\text{short}$ can be a SM singlet.  Clearly, such $\hpi$s cannot be pair-produced from Drell-Yan.  They cannot be pair-produced from a
resonant $\hrho$ decay either, because a resonant $\hrho$ couples to $\hpi$s in the same way as the SM gauge boson it mixes with does, but there is no
vertex with a SM gauge boson coupling to a singlet $\hpi$ pair. There are actually vertices with a singlet $\hpi$ and {\em two} gauge bosons, which can
allow a singlet $\hpi$ to be singly produced.  However, these vertices are analogous to the vertex for $\pi^0 \to \ga\ga$ in QED-QCD and contain a loop
factor $\sim 1/(16\pi^2)$.  Therefore, the production rate will be suppressed by $\sim 10^{-4}$, and we will not analyze their phenomenology in the rest
of this paper. (See Ref.\ \cite{VC} for constraints on such singlets.)

%%%%%%%%%%%%%%%%%%%%%%%%%%%%%%%%%%%%%%
\section{Benchmark Models and Results}
\Secl{benchmarks}
%%%%%%%%%%%%%%%%%%%%%%%%%%%%%%%%%%%%%%
The collider signatures of vectorlike confinement vary over a broad range depending on the SM charges carried by the hyperquarks. These charge assignments
determine the SM representations that the various $\hrho$ and $\hpi$ appear in. In this paper, our aim is to display the most characteristic features of
LHC phenomenology of vectorlike confinement. We thus choose to focus our attention on two benchmark models, their main difference being that the particles
in one benchmark carry QCD color and the other do not. The benchmark model with QCD color has large pair production cross section for colored $\hpi$. Such
$\hpi$ can be collider stable or decay to a pair of jets as explained in the previous section. The multi-jet final states have been studied in detail in
earlier work \cite{Coloron-1,Coloron-2}, and in this paper we will focus our attention on the phenomenology of R-hadrons, highlighting in particular that
vectorlike confinement gives rise to a unique final state of 4 R-hadrons in the same event. In contrast to the first benchmark model, the benchmark model
without QCD color has smaller pair production cross section for $\hpi$, which are color neutral in this case. On the other hand, such $\hpi$ are either
CHAMPs or decay to pairs of electroweak gauge bosons, therefore the final states are clean and backgrounds are much less of an issue compared to colored
final states. This section will be devoted to the detailed study of the final states in these two benchmark models. For completeness, the full underlying
definitions of the benchmark models are given in \App{fundamental}.

We should warn the reader at this point that the phenomenological lagrangians described below are merely tree-level parameterizations of production and
decay processes relevant for these final states, and should not be regarded as systematic low-energy effective field theories applicable at loop level.
While $\hpi$s are much lighter than the inverse of their size $\sim m_\hrho$ and can therefore be treated as point particles at energies $\ll m_\hrho$,
$\hrho$s are as heavy as the inverse of their size. Thus, there is no good field theoretical description of $\hrho$s. Fortunately, for the purposes of a
phenomenological study, we are only concerned with the resonant production of $\hrho$s and their subsequent decays to $\hpi$ pairs. Therefore it is
sufficient to model the production cross-section times branching ratios and how a single $\hrho$ propagates in the $s$-channel. The former can be
parameterized by introducing effective vertices for the production and decays, while the latter can be modeled by the standard Breit-Wigner form in the
narrow-width approximation. We are also leaving out many other bound states of hypercolor, which have masses $\sim m_\hrho$ unlike the $\hpi$, and cannot
be singly produced unlike the $\hrho$, and are therefore
much less relevant than $\hpi$ and $\hrho$.

%%%%%%%%%%%%%%%%%%%%%%%%%%%%%%%%%%%%%%%%%%%%%%%%%%%%%
\subsection{The Di-CHAMP/Multiphoton Benchmark Model}
\Secl{dichamp}
%%%%%%%%%%%%%%%%%%%%%%%%%%%%%%%%%%%%%%%%%%%%%%%%%%%%%
We introduce this benchmark in order to study the production of CHAMP pairs as well as multi-electroweak gauge bosons. The unique theoretical advantage of
this benchmark is that the analogy with QED-QCD mentioned above is correct not only qualitatively but also quantitatively (the number of colors and
flavors of the hypercolor sector are both 3), so the model is actually calculable by using the low-energy QED-QCD system as an analog computer. This
feature will be fully exploited below to reduce the number of adjustable parameters in the model.

The phenomenological highlights of this benchmark are resonant production of $\hrho$s (\Fig{rho-pi-pi}) that decay to (A) a pair of electrically charged,
collider-stable $\hpi$s (CHAMPs), or (B) a pair of $\hpi$s each of which subsequently decays to a pair of electroweak gauge bosons. In both cases, the
$\hpi$s can also be produced directly from an $s$-channel SM gauge boson (i.e.\ ``Drell-Yan'' production as in \Fig{Drell-Yan}). We will show that in
final state (A), the presence of the $\hrho$ resonance will be observable in the invariant mass distribution of CHAMP pairs. Final state (B) can also be
quite spectacular, especially when the final state is $3\ga+W$ (a $4\ga$ final state cannot be produced, as explained below). Since the four gauge bosons
come from the decays of two $\hpi$s originating from either a 2-to-2 Drell-Yan process or the decay of a $\hrho$ resonance, the signal rate is much larger
than SM four-gauge boson production, which is suppressed not only by the 2-to-4 phase space but also by a higher power of electroweak gauge couplings. We
will show how to reconstruct the full spectrum of the model and perform consistency checks to verify the expectations of vectorlike confinement in final
states (A) and (B).

Let us start by describing the low energy spectrum of the model relevant to the LHC phenomenology. (For the full detailed description of the model, see
\App{dichamp}.) Before electroweak symmetry breaking, there are three kinds of hyperpions, a weak doublet $\hpiD$, a weak triplet $\hpiT$, and a singlet
$\hpiS$. The $\SUC \otimes \SUL \otimes \UY$ quantum numbers of the $\hpiD$ and $\hpiT$ are $({\bf 1}, {\bf 2}, 3/2)$ and $({\bf 1}, {\bf 3}, 0)$,
respectively.%
\footnote{We normalize the $\UY$ charge such that the SM left-handed quark doublets have charge $1/6$.}
After electroweak symmetry breaking, it is more appropriate to distinguish their $\SUL$ components, $\hpiD \equiv (\hpiD^{++}, \hpiD^+)$ and $\hpiT \equiv
(-\hpiT^+, \hpiT^0, \hpiT^-)$ (as well as their antiparticles $\hpiD^{--} \equiv (\hpiD^{++})^\dag$, $\hpiD^{-} \equiv (\hpiD^{+})^\dag$, and $\hpiT^{-}
\equiv (\hpiT^{+})^\dag$). The $\hpiS$ is completely neutral under all SM gauge interactions, and is irrelevant for LHC physics as we explained at the end
of \Sec{pi-short}. The properties of the hyperpions in this benchmark are summarized in the following table:
\beq
\begin{tabular}{|c||c|c|c|c|}
\hline
                 & Color & Charge  & Mass                        & Decays to \\
\hline
$\hpiT^0$        & --    & $0$     & $m_{\rm T}$                 & $ZZ$, $Z\ga$, $\ga\ga$ \\
$\hpiT^\pm$      & --    & $\pm 1$ & $m_{\rm T} + \de m_{\rm T}$ & $W^\pm Z$, $W^\pm \ga$ \\
$\hpiD^\pm$      & --    & $\pm 1$ & $m_{\rm D}$                 & -- \\
$\hpiD^{\pm\pm}$ & --    & $\pm 2$ & $m_{\rm D} + \de m_{\rm D}$ & $\hpiD^\pm W^{\pm\ast}$ \\
$\hpiS$          & --    & $0$     & $m_{\rm S}$                 & $\ga\ga$, ($\ga Z$, $ZZ$) \\
\hline
\end{tabular}
\eeq
where the $W^{\pm*}$ stands for a SM fermion pair from an off-shell $W^\pm$,
and the $\pi_{\rm S}$ decay modes in the parentheses may not be kinematically allowed depending on the $\pi_{\rm S}$ mass.
The masses $m_{\rm D}$, $m_{\rm T}$ and $m_{\rm S}$ are parameterized by
\beq
  m_{\rm T}^2 &=& \fr{3 a m_\hrho^2}{16\pi^2} \cdot 2g_2^2
                 +6b m_\hrho m_2  \,,\nn\\
  m_{\rm D}^2 &=& \fr{3 a m_\hrho^2}{16\pi^2} \lt( \fr34 g_2^2 + \fr94 g_1^2 \rt)
                 +3b m_\hrho (m_2 + m_1)  \,,\nn\\
  m_{\rm S}^2 &=& 2b m_\hrho (m_2 + 2m_1)  \,,
\eql{di-champ:hpi-masses}
\eeq
where $m_1$ and $m_2$ are the masses of the hyperquarks $\psi_1$ and $\psi_2$ in the fundamental lagrangian (see \App{dichamp}), while $g_2$ and $g_1$ are
the SM $\SUL$ and $\UY$ gauge couplings, respectively. The coefficients $a$ and $b$ are positive $\cO(1)$ numbers, and can be determined by exploiting the
QED-QCD analogy of this model; specifically, $a$ can be extracted from $m_{\pi^\pm}^2 - m_{\pi^0}^2$ \cite{VC} while $b$ from $m_{\pi^0}$, and we find
\beq
   a = 1.2 \>,\quad b = 0.78 \text{ to } 1.6 \,,
\eql{quarkmassparam}
\eeq
where the factor-of-two uncertainty in $b$ is due to the uncertainty in $(m_u + m_d)/2$. Fortunately, this uncertainty will not enter our analysis below,
because we will parameterize the analysis directly in terms of $m_{\rm T}$ and $m_{\rm D}$.  The fact that $a$ and $b$ are $\cO(1)$ is sufficient for
checking that $m_{\rm S}$ is above the lower bound $\sim 100$ MeV discussed in Ref.\ \cite{VC}. The mass parameters $\de m_{\rm T}$ and $\de m_{\rm D}$
represents mass splitting due to electroweak symmetry breaking \cite{VC}, and their values are
\beq
  \de m_{\rm T} = 0.17\>\Gev \>,\quad \de m_{\rm D} = 1.1\>\Gev \,.
\eeq
These are very small and there is only one place where they matter; a $\hpiD^{\pm\pm}$ can now decay to a $\hpiD^{\pm}$ by emitting an off-shell $W^\pm$.%
\footnote{A $\hpiT^\pm$ can also decay to a $\hpiT^0$ by emitting an off-shell $W^\pm$, but the branching fraction of this mode is completely negligible to
$\hpiT^\pm \to W^\pm Z$, $W^\pm \ga$.}
The decay length of the $\hpiD^{\pm\pm}$ in this benchmark model is $\sim100~{\rm \mu m}$. Due to the small mass splitting, the off-shell $W^\pm$ can only
turn into an $e$-$\nu_e$/$\mu$-$\nu_\mu$ pair or a single meson. Unfortunately, the mesons and leptons from $\hpiD^{\pm\pm}$ decays will not be observable
at the LHC because they are too soft, only carrying 1 GeV of energy/momentum. Therefore, once produced, a $\hpiD^{\pm\pm}$ behaves just like a $\hpiD^\pm$
for all practical purposes. The $\hpiD^\pm$ is collider stable, thus a CHAMP. (For the eventual decay of $\hpiD^\pm$, see \App{dichamp}.)

Next, the $\hrho$s relevant to the LHC phenomenology are $W'^{\pm}$, $W'^{3}$, and $B'$, which behave as heavy versions of $W^{\pm}$, $W^3$ (the 3rd
component of the $\SUL$ gauge field), and $B$ (the $\UY$ gauge field). Unlike the SM $W^3$ and $B$, $W'^{3}$ and $B'$ already have $\cO(\text{TeV})$
masses without electroweak symmetry breaking, so their mixing due to electroweak symmetry breaking are inconsequential. Corrections to $\hrho$ masses due
to hyperquark masses and SM gauge interactions are small, so we will use a single mass $m_\hrho$ for all three. As their names suggest, $W'^{\pm}_\mu$,
$W'^{3}_\mu$, and $B'_\mu$ couple to SM fermions in the same way as $W^\pm$, $W^3$, and $B$ do, except that their coupling constants are different from
the SM values. It is through these couplings that they can be resonantly produced at the LHC from a $q$-$\bar{q}$ initial state.

The effective $q$-$\bar{q}$-$\hrho$ vertices parameterizing the resonant $\hrho$ production are given by
\beq
  \cL_\text{int}
  &=& +\fr{g_2^2\de}{2\sqrt2} W'^{+}_\mu \bar{u} \ga^\mu (1-\ga_5) d
      +\fr{g_2^2\de}{2\sqrt2} W'^{-}_\mu \bar{d} \ga^\mu (1-\ga_5) u  \nn\\
   && +\fr{g_2^2\de}{4} W'^{3}_\mu \lt[ \bar{u} \ga^\mu (1-\ga_5) u - \bar{d} \ga^\mu (1-\ga_5) d \rt]  \nn\\
   && +\fr{g_1^2\de\sqrt3}{12} B'_\mu \lt[ \bar{u} (5+3\ga_5) u - \bar{d} (1+3\ga_5) d \rt]  ,
\eql{hrho-q-qbar:dichamp}
\eeq
where $g_2$ and $g_1$ are the SM $\SUL$ and $\UY$ gauge couplings, respectively.%
\footnote{Our sign convention is such that the SM coupling of, e.g., a $W^+$ to $u$ and $d$ quarks (ignoring the CKM mixings) is given by
$\displaystyle{\cL_\text{int} = -\fr{g_2}{2\sqrt2} W^+_\mu \bar{u} \ga^\mu (1-\ga_5) d}$, and the metric is taken to be ``mostly minus'', i.e.\
$(g_{\mu\nu}) = \text{diag}\,(+1,-1,-1,-1)$. $\ga_5$ is $+1$ and $-1$ for right-handed and left-handed fermions, respectively.}
There are similar couplings for the 2nd and 3rd generations as well, with the same strengths. (We ignore CKM mixing in this paper.) The $\de$ parameter
can be extracted from the $e^+ e^- \to \rho$ rate, by exploiting the exact analogy between this benchmark model and the QED-QCD system:
\beq
  \de = 0.20  \,.
\eql{delta}
\eeq
Once produced, $W'^{\pm}$, $W'^{3}$, and $B'$ decay to $\hpiD$ or $\hpiT$ pairs. The $\hpiD$ pair leads to the di-CHAMP signal, while the $\hpiT$ pair
leads to the multiple gauge boson signal. We will next discuss these two signals in detail. In our study, we focus on three mass points, given in the
table below. In \Fig{pipairxsec} we display the production cross-section for the $\hpi$s as a function of $m_{\hpi}$.
To be conservative, for each $m_\hpi$, $m_{\hrho}$ is taken at its maximal value, corresponding zero hyperquark masses.
Finite hyperquark masses are expected to lower the value of $m_{\hrho}$ for a given $m_\hpi$ and thus
increase the production cross-sections.
\beq
\begin{tabular}{|c||c|c|c|}
\hline
mass point & $m_{\hrho}$ (TeV)  & $m_{\rm D}$ (GeV)  & $m_{\rm T}$ (GeV)\\
\hline
1          & $1.5$              & $300$              & $300$ \\
2          & $2.5$              & $300$              & $355$ \\
3          & $2.5$              & $600$              & $600$ \\
\hline
\end{tabular}
\eql{dichamp_points}
\eeq

\DOUBLEFIGURE[t]{plot_pipairproductionxsec.eps, width=1.0\linewidth} {plot_CHAMP_shat.eps, width=1.0\linewidth} {\figl{pipairxsec}The pair-production
cross-sections for all $\hpi$s at the LHC ($E_{\rm CM} = 14$ TeV). Hyperquark masses are taken to be zero for this plot. Note that a single value for
$m_\hpi$ corresponds to different values of $m_\hrho$ for $\hpiD$ and $\hpiT$.} {\figl{champ:invmass}The differential CHAMP pair-production cross-sections
at the LHC ($E_{\rm CM}=14$ TeV) for the three mass points used in our analysis.}

%%%%%%%%%%%%%%%%%%%%%%%%%%%%%%%%%%%
\subsubsection{The Di-CHAMP Signal}
%%%%%%%%%%%%%%%%%%%%%%%%%%%%%%%%%%%
This section focuses on the kinematic features of the $\hpiD$-pair final states as well as issues of triggering. Since the $\hpiD^\pm$ is collider-stable
and charged (but color-neutral), it will appear in the LHC detector like a massive ``muon''. Recall that $\hpiD^{\pm\pm}$ pair production will be
indistinguishable from $\hpiD^{\pm}$ pair production because the decay $\hpiD^{\pm\pm} \to \hpiD^{\pm} + W^{\pm\ast}$ is prompt and unobservable due to
the small mass splitting. Therefore the two cross-sections can be added to give the total ``CHAMP pair production'' cross-section, which is plotted in
\Fig{pipairxsec}.

The production of $\hpiD$ through SM Drell-Yan can be described by
\beq
  \cL_\text{int}
  &=& -\fr{ig_2}{\sqrt2} W^{+\mu} \lt[ \hpiD^{--} (\del_\mu \hpiD^+) - (\del_\mu \hpiD^{--}) \hpiD^+ \rt]
      -\fr{ig_2}{\sqrt2} W^{-\mu} \lt[ \hpiD^- (\del_\mu \hpiD^{++}) - (\del_\mu \hpiD^-) \hpiD^{++} \rt]  \nn\\
   && -ig_{Z}^{++} Z^\mu \lt[ \hpiD^{--} (\del_\mu \hpiD^{++}) - (\del_\mu \hpiD^{--}) \hpiD^{++} \rt]
      -ig_{Z}^{+} Z^\mu \lt[ \hpiD^{-} (\del_\mu \hpiD^{+}) - (\del_\mu \hpiD^{-}) \hpiD^{+} \rt]  \nn\\
   && -2ie A^\mu \lt[ \hpiD^{--} (\del_\mu \hpiD^{++}) - (\del_\mu \hpiD^{--}) \hpiD^{++} \rt]
      -ie A^\mu \lt[ \hpiD^{-} (\del_\mu \hpiD^{+}) - (\del_\mu \hpiD^{-}) \hpiD^{+} \rt]  \,,
\eeq
where $e$ is the $\U(1)_\text{EM}$ gauge coupling, and
\beq
   g_Z^{++} = \fr{g_2}{\cos\tht} \lt( \fr12 - 2\sin^2\!\tht \rt)  \>,\quad
   g_Z^{+}  = \fr{g_2}{\cos\tht} \lt( -\fr12 - \sin^2\!\tht \rt)  \,,
\eeq
$\tht$ being the weak mixing angle. There exist also four-point vertices with two gauge bosons and two $\hpiD$s, but we omit them because $\hpiD$
pair-production through electroweak gauge boson fusion is negligible compared to the processes of Figures \fig{rho-pi-pi} and \fig{Drell-Yan}.

\DOUBLEFIGURE[t]{plot_CHAMP_pt.eps, width=1.0\linewidth} {plot_CHAMP_eta.eps, width=1.0\linewidth} {\figl{champ:pt}The $p_{\rm T}$ distribution of CHAMP
pairs for the three mass points.} {\figl{champ:eta}The rapidity distribution of CHAMP pairs for the three mass points.}

The production of $\hpiD$s from the decays of $\hrho$s is described by:
\beq
  \cL_\text{int}
  &=& -\fr{ig_\hrho}{\sqrt2} W'^{+\mu} \lt[ \hpiD^{--} (\del_\mu \hpiD^+) - (\del_\mu \hpiD^{--}) \hpiD^+ \rt]
      -\fr{ig_\hrho}{\sqrt2} W'^{-\mu} \lt[ \hpiD^- (\del_\mu \hpiD^{++}) - (\del_\mu \hpiD^-) \hpiD^{++} \rt]
       \nn\\
   && -\fr{ig_\hrho}{2} W'^{3\mu} \lt[ \hpiD^{--} (\del_\mu \hpiD^{++}) - (\del_\mu \hpiD^{--}) \hpiD^{++} \rt]
      +\fr{ig_\hrho}{2} W'^{3\mu} \lt[ \hpiD^{-} (\del_\mu \hpiD^{+}) - (\del_\mu \hpiD^{-}) \hpiD^{+} \rt]
       \nn\\
   && -\fr{ig_\hrho\sqrt3}{2} B'^{\mu}
       \lt[ \hpiD^{--} (\del_\mu \hpiD^{++}) - (\del_\mu \hpiD^{--}) \hpiD^{++}
           +\hpiD^{-} (\del_\mu \hpiD^{+}) - (\del_\mu \hpiD^{-}) \hpiD^{+}
       \rt]  \,,
\eeq
combined with \eq{hrho-q-qbar:dichamp} which describes the resonant production of the $\hrho$s. The coupling constant $g_\hrho$ can be extracted using the
QED-QCD analogy to be
\beq
  g_\hrho = 6.0  \,.
\eql{ghrho}
\eeq

We have implemented the above effective vertices into CALCHEP 2.5.4 \cite{calchep}. For all Monte Carlo analyses in this paper we use CTEQ6 PDF's
\cite{Pumplin:2002vw}. For the three values of ($m_{\rm D}$, $m_\hrho$) in \eq{dichamp_points}, the invariant mass distribution of the CHAMP pairs is
shown in \Fig{champ:invmass} at the LHC with $14~{\rm TeV}$ CM energy. The point of this plot is to highlight a large deviation from the pure Drell-Yan
distribution due to the $\hrho$ resonance. The mass point 2 assumes vanishing hyperquark masses, and is therefore the most conservative choice,
corresponding to the maximum possible mass gap between the CHAMP and the $\hrho$. But even in this case there is a significant deviation in the invariant
mass distribution from pure Drell-Yan. The $m_{\rm D}$-$m_\hrho$ gap is narrower for mass points 1 and 3 due to nonzero hyper-quark masses, in which case
the existence of the resonance becomes quite pronounced.

Next, we show the $p_{\rm T}$ distribution of the CHAMPs in \Fig{champ:pt} and their rapidity distribution in \Fig{champ:eta}.
The most important information in
these plots is the fact that both CHAMPs are produced with high $p_{\rm T}$ and within the detector acceptance most of the time.
In particular, using the ATLAS muon coverage
parameters ($|\eta|<2.5$), both CHAMPs will be within acceptance with an efficiency of $0.94$ for mass point 1, $0.91$ for mass point 2 and $0.98$ for
mass point 3. Note that in both Drell-Yan and $\hrho$ resonance production, the CHAMP pair originates from an intermediate state with spin-1, so the
angular distribution corresponds to $\cos\tht$ in the center-of-momentum frame of the CHAMP pair. This is shown in \Fig{champ:thetaCM}.

\DOUBLEFIGURE[t]{plot_CHAMP_thetaCM.eps, width=1.0\linewidth} {plot_CHAMP_beta.eps, width=1.0\linewidth} {\figl{champ:thetaCM}For CHAMP pair production,
the production angle in the CM frame is plotted. This can be used as conclusive evidence of a spin-1 intermediate state in the s-channel.}
{\figl{champ:beta}The velocity distribution of CHAMP pairs for the three mass points. Due to the spin-1 intermediate state, the distribution is peaked
away from threshold.}

As we mentioned before, the $s$-channel spin-1 intermediate state forces the CHAMPs to be produced away from threshold, as can be seen in the velocity
distribution of the CHAMPs in \Fig{champ:beta}. This also addresses a significant worry in detecting the CHAMPs, namely the issue of triggering. The
CHAMPs will be triggered on when they reach the muon chamber, but since they are massive and have low speeds, they might be labelled as out-of-sync events
and thus thrown away.  However, since we have demonstrated that the CHAMPs are typically produced with $\beta\sim 1$, this worry is alleviated.
To check this quantitatively, let us look at
the time-lag which is defined as the additional time required for a particle to reach the muon chamber relative
to a massless particle that was produced in the same bunch crossing. We use the parameters for the ATLAS muon system (which is more conservative in terms
of triggering as the distances are larger) and we differentiate the barrel region ($|\eta|<1.4$) from the endcap region ($1.4<|\eta|<2.5$). For the barrel
region we calculate the time to get to a radius of $7.5~{\rm meters}$ from the crossing region, and for the endcap we calculate the time to get to
$|z|=14.5~{\rm meters}$. In Figures \fig{champ:tof1} and \fig{champ:tof2} we plot the time-lag for the earlier and later CHAMP in the event, respectively.
As pointed out in Ref.\ \cite{CHAMPs-Rhadrons}, for a time-lag below $25$ ns, triggering should be efficient, so \Fig{champ:tof1} and \Fig{champ:tof2}
look very encouraging.

An experimental technique to separate slow vs.\ fast CHAMPs (and vs.\ muons, of course) was recently proposed in Ref.\ \cite{Todd-Jay}. They define ``slow
CHAMPs'' as those with $0.6< \be < 0.8$ while ``fast CHAMPs'' as those with $\be > 0.95$. Slow CHAMPs can be identified by conventional $dE/dx$ and
time-of-flight (TOF) measurements. On the other hand, they showed that fast CHAMPs can be distinguished from muons by using the fact that muons with such
high energy ($E > 317$ GeV in copper and $E > 581$ GeV in silicon) will actually lose energy by bremsstrahlung while CHAMPs behave as minimum ionizing
particles. In fact, according to their analysis, it is even possible that fast CHAMPs are the ones the LHC can recognize first in the early data. Since
the ratio of slow to fast CHAMPs reflects the $m_\hrho/m_{\rm D}$ ratio, it can be used as a consistency check for $m_\hrho/m_{\rm D}$ extracted by
fitting, e.g., the invariant mass distribution.

\DOUBLEFIGURE[t]{plot_CHAMP_tofearly.eps, width=1.0\linewidth} {plot_CHAMP_toflate.eps, width=1.0\linewidth} {\figl{champ:tof1}The time-lag of the first
CHAMP to arrive at the muon system (further details in the text).} {\figl{champ:tof2}The time-lag of the second CHAMP to arrive at the muon system
(further details in the text).}

%%%%%%%%%%%%%%%%%%%%%%%%%%%%%%%%%%%%%%%%%%%%
\subsubsection{The $3\gamma + W^\pm$ Signal}
%%%%%%%%%%%%%%%%%%%%%%%%%%%%%%%%%%%%%%%%%%%%
This section focuses on the study of multi-photon final states from the production and decay of $\hpiT$s. The $\hpiT$ pair production through Drell-Yan
processes are described by
\beq
  \cL_\text{int}
  &=& +ig_2 W^{+\mu} \lt[ \hpiT^- (\del_\mu \hpiT^0) - (\del_\mu \hpiT^-) \hpiT^0 \rt]
      +ig_2 W^{-\mu} \lt[ \hpiT^0 (\del_\mu \hpiT^+) - (\del_\mu \hpiT^0) \hpiT^+ \rt]  \nn\\
   && -ig_2\cos\tht \, Z^\mu \lt[ \hpiT^{-} (\del_\mu \hpiT^{+}) - (\del_\mu \hpiT^{-}) \hpiT^{+} \rt]  \nn\\
   && -ie A^\mu \lt[ \hpiT^{-} (\del_\mu \hpiT^{+}) - (\del_\mu \hpiT^{-}) \hpiT^{+} \rt]  \,.
\eeq
Again, we omitted four-point vertices with two gauge bosons and two $\hpiT$s as those are irrelevant at the LHC.
The $\hpiT$ pair production from $\hrho$ are described by \eq{hrho-q-qbar:dichamp} for the production of the $\hrho$ and
\beq
  \cL_\text{int}
  &=& +ig_\hrho W'^{+\mu} \lt[ \hpiT^- (\del_\mu \hpiT^0) - (\del_\mu \hpiT^-) \hpiT^0 \rt]
      +ig_\hrho W'^{-\mu} \lt[ \hpiT^0 (\del_\mu \hpiT^+) - (\del_\mu \hpiT^0) \hpiT^+ \rt]  \nn\\
   && -ig_\hrho W'^{3\mu} \lt[ \hpiT^{-} (\del_\mu \hpiT^{+}) - (\del_\mu \hpiT^{-}) \hpiT^{+} \rt]  \,,
\eeq
for the decays (where $g_\hrho$ is given in \eq{ghrho}). Note that there is no way to pair-produce $\hpiT^0$s, neither via Drell-Yan nor from $\hrho$
decay. This can be also understood in terms of angular momentum conservation and the $\hpiT^0$'s Bose statistics. We refer the reader back to
\Fig{pipairxsec} for the production cross-sections of various $\hpiT$ pairs.

Next, we turn our attention to the decay branching fractions of the $\hpiT$. Unlike $\hpiD$s, which are collider stable, $\hpiT$s decay promptly to SM
electroweak gauge boson pairs, which are analogous to $\pi^0 \to \ga\ga$ in QED-QCD. Like $\pi^0 \to \ga\ga$, the vertices for these decays are strictly
determined by anomalies, and are given by
\beq
 \cL_\text{int}
 &=& +\fr{3g_1 g_2 \sin\tht}{8\pi^2 f_\hpi} \ep^{\mu\nu\rho\sg} \,
      \hpiT^+ (\del_\mu W^-_\nu) (\del_\rho Z_\sg)
     -\fr{3g_1 g_2 \cos\tht}{8\pi^2 f_\hpi} \ep^{\mu\nu\rho\sg} \,
      \hpiT^+ (\del_\mu W^-_\nu) (\del_\rho A_\sg)  \nn\\
  && +\fr{3g_1 g_2 \sin\tht}{8\pi^2 f_\hpi} \ep^{\mu\nu\rho\sg} \,
      \hpiT^- (\del_\mu W^+_\nu) (\del_\rho Z_\sg)
     -\fr{3g_1 g_2 \cos\tht}{8\pi^2 f_\hpi} \ep^{\mu\nu\rho\sg} \,
      \hpiT^- (\del_\mu W^+_\nu) (\del_\rho A_\sg)  \nn\\
  && +\fr{3g_1 g_2 \sin 2\tht}{16\pi^2 f_\hpi} \ep^{\mu\nu\rho\sg} \,
      \hpiT^0 (\del_\mu Z_\nu) (\del_\rho Z_\sg)
     -\fr{3g_1 g_2 \cos 2\tht}{8\pi^2 f_\hpi} \ep^{\mu\nu\rho\sg} \,
      \hpiT^0 (\del_\mu Z_\nu) (\del_\rho A_\sg)  \nn\\
  && -\fr{3g_1 g_2 \sin 2\tht}{16\pi^2 f_\hpi} \ep^{\mu\nu\rho\sg} \,
      \hpiT^0 (\del_\mu A_\nu) (\del_\rho A_\sg)  \,,
\eeq
where $\ep^{\mu\nu\rho\sg}$ is a totally anti-symmetric tensor with $\ep^{0123} \equiv +1$.%
\footnote{Because of the non-Abelian nature of $\SUL$, there are also decays to 3 and 4 gauge bosons as well. However, their rates are suppressed by
multi-body phase space factors, and we will only dwell on the dominant decay modes in this paper.}
The branching fractions following from these vertices are shown in \Fig{BFpiplus} and \Fig{BFpi0}. Using the QED-QCD analogy, the hyperpion decay constant
$f_\hpi$ can be obtained just by scaling up the pion decay constant $f_\pi$, i.e., $f_\hpi = f_\pi m_\hrho/m_\rho$, and is $\cO(100)$ GeV. Note that the
widths of $\hpiT$s are extremely narrow, so their reconstructed widths will be dominated by the detector resolution.
The smallness of these couplings
and the absence of tree
level couplings of hyperpions to SM fermions means that the $\hpiT$ are practically never resonantly produced, thus they are not constrained by di-boson
resonance searches or fermiophobic Higgs searches at LEP (this is discussed in more detail in Ref.\ \cite{VC}).

\DOUBLEFIGURE[t]{plot_piplusBR.eps, width=1.0\linewidth} {plot_pi0BR.eps, width=1.0\linewidth} {\figl{BFpiplus}Branching fractions of $\hpiT^{\pm}$.}
{\figl{BFpi0}Branching fractions of $\hpiT^{0}$.}

Let us now study the signals of $\hpiT$ pair production at the LHC. Among possible final states, the one with the maximum number of photons is the
$3\ga+W^\pm$ final state from the production and decay of $\hpiT^0$-$\hpiT^\pm$. Note that this particular decay channel is in fact the one with the
largest branching fraction (albeit not by a large margin), so in our study we will focus on this channel exclusively.
Final states with two photons $WZ+\ga\ga$ and $W\ga+Z\ga$, while still interesting, present additional difficulties compared to the $3\ga$-$W$ final
state due to non-resonant photons (in the latter case), small leptonic branching fractions and combinatoric issues.
Certain decay channels can look like a heavy SM Higgs decaying to $W$s and $Z$s,
but note that the number of events with two or three photons will be {\em comparable,} clearly distinguishing a $\hpiT$ from the SM Higgs.

The backgrounds for multiple photons can be divided into ``real'' and ``fake'' ones, the former being processes in which all photons are produced from an
actual SM process, while in the latter one or more of the photon signatures is faked by an electron or a jet. While publicly available matrix element
Monte Carlo generators are capable of generating background SM processes with ``real'' photons, the study of ``fake'' backgrounds necessitates the usage
of a sophisticated detector simulation such as GEANT, to which we do not have access. However, we can take some guidance from collider studies of diphoton
final states, in particular searches of a light SM Higgs boson \cite{ATLAS-H2gaga, CMS-H2gaga}, where after applying analysis cuts the fake backgrounds
can be brought down to the level of irreducible backgrounds, up to factors of order one. Therefore, in our preliminary study, we will scale up the
irreducible background by a factor of 10 in order to mimic backgrounds with fakes and leave it to experimentalists to perform a study with more
sophisticated backgrounds using a full detector simulation.

\DOUBLEFIGURE[t]{plot_3gamma_efficiency.eps, width=1.0\linewidth} {plot_3gamma_mass_b1.eps, width=1.0\linewidth} {\figl{3gamma:efficiency}For mass points
1-3 we plot the signal efficiency as a function of the $p_{\rm T}$ cut on all three photons. Only photons with $|\eta|<2.5$ are considered.}
{\figl{3gamma:mass1}For mass point 1 we plot the invariant mass of all possible photon pairs for signal and background. The $\hpiT^0$ resonance is clearly
visible on top of the smooth SM background as well as combinatoric background from the signal.}

We generate signal events using CALCHEP 2.5.4 \cite{calchep} as in our study of the CHAMP final state, but we also pass the parton level events through
Pythia \cite{pythia} for initial and final state showering and hadronization and then through PGS4 \cite{PGS} for energy smearing and detector effects,
where we use the CMS parameter set. For backgrounds we use the matching utility in MadEvent \cite{Madevent} to combine $3\ga$(+jet(s)) processes (MLM
matching with ickkw set to 1), and also pass these events through Pythia and PGS. We scale up the cross-sections for the background by a factor of 10 as
mentioned above.

Since in SM processes photons are usually emitted in radiative processes, backgrounds can be very efficiently reduced by demanding high-$p_{\rm T}$ central
photons. In \Fig{3gamma:efficiency} we plot the fraction of signal events that have three photons within $|\eta|<2.5$ as a function of a $p_{\rm T}$ cut
(applied on all three photons). In the remainder of this analysis, for both signal and background we will select events with three photons, all having
$p_{\rm T} > 120$ GeV as well as $|\eta|<2.5$.

For all such events, we then proceed to reconstruct the $\hpiT^0$, from the decay of which two of the three photons originate. We calculate the invariant
mass of all three possible photon pairings for signal and background, and plot the obtained values. The results are displayed in \Fig{3gamma:mass1}
through \Fig{3gamma:mass3} for mass points 1 through 3. The presence of the $\hpiT^0$ resonance is unmistakable on top of the SM background as well as the
combinatoric background from signal, both of which are smooth over the region of the resonance peak. We then take over the central value for the resonance
as the hyperpion mass and use that value to reconstruct the $\hpiT^{\pm}$. We also assign the photon pair that gives the closest invariant mass to the
central value as originating from the $\hpiT^0$ and unpaired photon as originating from the $\hpiT^{\pm}$. Also, we can now use the $\hpiT^0$ resonance to
further reduce the background (which was already very small) by only looking at events that have a diphoton resonance at the correct mass. Thus,
background is expected to be negligible for the reconstruction of the $\hpiT^{\pm}$, and the rest of our analysis will be signal-only.

\DOUBLEFIGURE[t]{plot_3gamma_mass_b2.eps, width=1.0\linewidth} {plot_3gamma_mass_b3.eps, width=1.0\linewidth} {\figl{3gamma:mass2}For mass point 2 we plot
the invariant mass of all possible photon pairs for signal and background. The $\hpiT^0$ resonance is clearly visible on top of the smooth SM background
as well as combinatoric background from the signal.} {\figl{3gamma:mass3}For mass point 3 we plot the invariant mass of all possible photon pairs for
signal and background. The $\hpiT^0$ resonance is clearly visible on top of the smooth SM background as well as combinatoric background from the signal.}

For the reconstruction of $\hpiT^{\pm}$, we take the unpaired photon and combine it with the best $W^{\pm}$ candidate which we define as follows. In
leptonic $W$ decays ($e$ or $\mu$ only) we assume that the missing $E_{\rm T}$ is entirely due to a neutrino, and solve for its rapidity by demanding that
it reconstruct an on-shell $W$. If there are no solutions, the event is discarded. If there are multiple solutions, we take the one for which the $W$
candidate together with the unpaired photon gives the closest invariant mass to that obtained from the $\hpiT^0$ resonance. For hadronic $W$ decays, we
use the fact that the $W$ from the $\hpiT^{\pm}$ decay is moderately boosted. We take all pairs of jets ($p_{\rm T}>20$ GeV) in the event and look for a
pairing with $\Delta R_{\rm jj}<2.0$ which reconstructs an invariant mass of 70 GeV $<m_{\rm jj}<90$ GeV. If there is exactly one such pairing, we take
the total 4-momentum of the pair to be the $W$, in all other cases we discard the event. We then combine the leptonic and hadronic $W$-decay events that
satisfy these requirements and calculate the invariant mass of the reconstructed $W$ together with the unpaired photon. The results for mass points 1
through 3 are displayed in \Fig{3gamma:reconpimass1} through \Fig{3gamma:reconpimass3}, and agree very well with the true mass of the $\hpiT^{\pm}$.

\DOUBLEFIGURE[t]{plot_3gamma_reconpimass_b1.eps, width=1.0\linewidth} {plot_3gamma_reconpimass_b2.eps, width=1.0\linewidth} {\figl{3gamma:reconpimass1}For
mass point 1 we plot the mass of the reconstructed $\hpiT^{\pm}$ as explained in detail in the text. This plot is signal-only.}
{\figl{3gamma:reconpimass2}For mass point 2 we plot the mass of the reconstructed $\hpiT^{\pm}$ as explained in detail in the text. This plot is
signal-only.}

For the final step in the analysis, we combine the 4-momenta of the reconstructed $\hpi^0$ and the reconstructed $\hpi^{\pm}$ (only in events that satisfy
the requirements of the reconstruction procedure outlined above) to look for the contribution from the resonant $\hrho$ production and its subsequent
decay. The results for the three mass points are shown in \Fig{3gamma:reconrhomass},
exhibiting a bump or shoulder at high energy similar to \Fig{champ:invmass}, reflecting the $\hrho$ resonance.
Combined with the results of the CHAMP analysis, this provides a strong consistency check of the underlying theory of vectorlike
confinement. From the masses of $\hpiT$, $\hpiD$ and $m_\hrho$, one can test the relations \eq{di-champ:hpi-masses} or similar relations in other
variations of vectorlike confinement to probe the underlying hyperflavor symmetry structure.

%%%%%%%%%%%%%%%%%%%%%%%%%%%%%%%%%%%%%%
\subsection{The Di-R-hadron Benchmark}
%%%%%%%%%%%%%%%%%%%%%%%%%%%%%%%%%%%%%%
This benchmark model contains long lived $\hpi$ that carry color as well as $\hpi$ that decay promptly to gluon pairs. The phenomenology of the latter has
been extensively studied in Refs.\ \cite{Coloron-1} and \cite{Coloron-2}, so we will devote ourselves in this paper to studying the signals of the former.
Such hyperpions will promptly hadronize with quarks and gluons, forming massive ($\sim$ a few hundred GeV--1 TeV) stable QCD bound states --- R-hadrons.
Like CHAMPs in our earlier benchmark model, the R-hadrons can be pair-produced from the decay of a resonant (color octet) $\hrho$ (as in \Fig{rho-pi-pi})
as well as through an $s$-channel gluon (as in \Fig{Drell-Yan}) for $q$-$\bar{q}$ initial states. For a $g$-$g$ initial state, there exist additional
diagrams as illustrated in \Fig{gg-to-pipi}.

As described in the full detail in \App{Rhadron}, this benchmark model is analogous to QCD with 3 colors and 4 light flavors.  Therefore, unlike in the
di-CHAMP benchmark model, which is exactly analogous to the 3-color-3-flavor QCD of the real world, the parameters in this benchmark model cannot be
determined as precisely by using the analogy with QED/QCD.  However, since the difference is only 3 vs.\ 4 flavors, we expect that the QED/QCD analogy
should still provide good estimates.\footnote{The one place where the QED-QCD analogy completely fails is the process, $gg \to \hrho$, which has no analog
in QED-QCD because of the Abelian nature of the $U(1)_\text{EM}$. For this process we employ the estimate given in Refs.\ \cite{Coloron-1, Coloron-2},
which we expect to hold within an $\cO(1)$ uncertainty.} Therefore, we employ the 3-color-3-flavor values for incalculable parameters in our computations
below, and we will only focus on robust conclusions that are insensitive to $\cO(1)$ uncertainties.  Such a rough strategy is possible because the
production rates are large, as we will see below. The large signal rates also allow us to ignore the details of how the colored long-lived $\hpi$s
hadronize into R-hadrons and interact with the detector material. Even if such effects render an $\cO(1)$ fraction of the signal events unobservable, the
rates are large enough such that reconstruction is still straightforward. We direct the interested reader to Ref.\ \cite{CHAMPs-Rhadrons} for a detailed
analysis of R-hadrons in collider detectors.

\DOUBLEFIGURE[t]{plot_3gamma_reconpimass_b3.eps, width=1.0\linewidth} {plot_3gamma_reconrhomass.eps, width=1.0\linewidth} {\figl{3gamma:reconpimass3}For
mass point 3 we plot the mass of the reconstructed $\hpiT^{\pm}$ as explained in detail in the text. This plot is signal-only.}
{\figl{3gamma:reconrhomass}For mass points 1 through 3 we plot the mass of the reconstructed $\hpiT^{0}$-$\hpiT^{\pm}$ system as explained in detail in
the text. This plot is signal-only.}

Let us start by describing the low energy spectrum of the model relevant to the LHC phenomenology. (For the full detailed description of the model, see
\App{Rhadron}.) The hyperpions come in a color octet and triplet, $\hpi_8$ and $\hpi_3$, with the $\SUC \otimes \SUL \otimes \UY$ quantum numbers $({\bf
8}, {\bf 1}, 0)$ and $({\bf 3}, {\bf 1}, -4/3)$, respectively.  Note that they are both weak singlets. There is also a completely SM singlet hyperpion,
$\hpi_1$, which can be resonantly produced from a $g$-$g$ initial state at the LHC, and has a rare decay mode (with branching fraction of ${\mathcal
O}(1\%)$) to a pair of photons. Since the $\hpi_1$ is very light, constraints on its existence from diphoton resonance searches at the Tevatron were
discussed in Ref.\ \cite{VC}, and its mass was found to be unconstrained. Since the $\hpi_1$ is produced near threshold, the photons from its decay rarely
pass the cuts of the search. While very difficult, it may be possible to search for it at the LHC. We will however limit ourselves here to
the study of the R-hadrons in this benchmark model which can be looked for much more easily.

The properties of the hyperpions are summarized in the following table:
\beq
\begin{tabular}{|c||c|c|c|c|}
\hline
         & Color     & Charge & Mass  & Decays to \\
\hline
$\hpi_8$ & ${\bf 8}$ & $0$    & $m_{\hpi_8}$ & $gg\ \gg\ gZ$, $g\ga$ \\
$\hpi_3$ & ${\bf 3}$ & $-4/3$ & $m_{\hpi_3}$ & collider stable \\
$\hpi_1$ & ${\bf 1}$ & $0$    & $m_{\hpi_1}$ & $gg\ \gg\ \ga\ga\ \gg\ \ga Z$, $ZZ$ \\
\hline
\end{tabular}
\eeq
The mass parameters $m_{\hpi_8}$, $m_{\hpi_3}$ and $m_{\hpi_1}$ are given by
\beq
  m_{\hpi_8}^2 &=& \fr{3 a m_\hrho^2}{16\pi^2} \cdot 3g_3^2
                  +6b m_\hrho m_3  \,,\nn\\
  m_{\hpi_3}^2 &=& \fr{3 a m_\hrho^2}{16\pi^2} \lt( \fr43 g_3^2 + \fr{16}{9} g_1^2 \rt)
                  +3b m_\hrho (m_3 + m_1)  \,,\nn\\
  m_{\hpi_1}^2 &=& \fr{3b}{2} m_\hrho (m_3 + 3m_1)  \,,
\eql{R-hadron:hpi-masses}
\eeq
where $m_1$ and $m_3$ are the masses of the hyperquarks $\psi_1$ and $\psi_3$ in the fundamental theory (see \App{Rhadron} for details).  As explained
above, we cannot determined $a$ and $b$ and we employ the values \eq{quarkmassparam} as estimates. Fortunately, this uncertainty will not affect our
analysis below, as we will only concern ourselves with $\hpi_8$ and $\hpi_3$, so we can parameterize physics in terms of $m_{\hpi_8}$ and $m_{\hpi_3}$
rather than $m_3$ and $m_1$.

\EPSFIGURE[t]{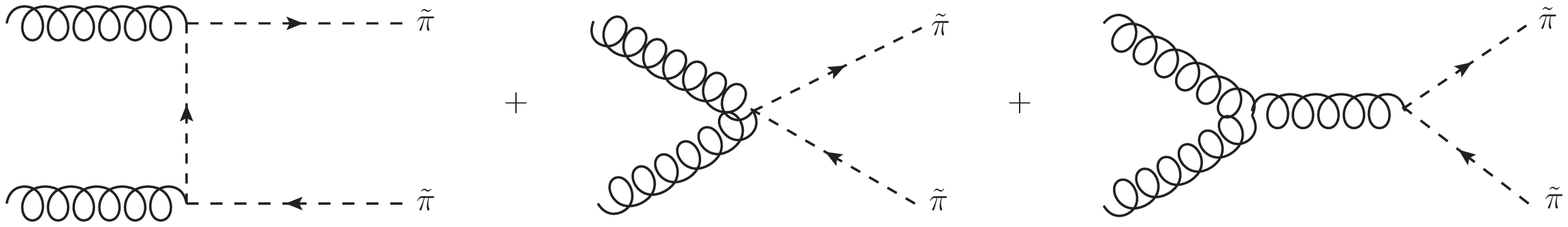, width=1.0\linewidth}
{\figl{gg-to-pipi}The diagrams leading to the pair production of colored $\hpi$ from a $g$-$g$ initial state.}

We will denote the $\hrho$s that can mix with the SM gauge bosons as $g'$ and $B'$, which behave as heavy versions of the gluon $g$ and the $\UY$ gauge
boson $B$, respectively. As already mentioned, corrections to the hyper-rho masses due to the hyperquark masses and SM gauge interactions are small, so
both hyper-rho masses can simply be denoted by a single number, $m_\hrho$. No other hypermesons or hyperbaryons can be resonantly produced,
and since they are heavy ($\sim m_\rho$) their pair-production cross-sections are much smaller than those of the $\hpi$s, which are much lighter.

The effective vertices parameterizing the resonant $q\bar{q} \to g', B'$ processes (as in \Fig{rho-pi-pi}) are given by%
\footnote{Our sign convention is such that the SM couplings of the gluon $G_\mu$ to quarks are given by $\displaystyle{\cL_\text{int} = -\fr{g_3}{2}
G^a_\mu \, \bar{q} \ga^\mu \la^a q}$.}
\beq
  \cL_\text{int}
  = +\fr{g_3^2\de}{2} G'^{a}_\mu \lt[ \bar{u} \ga^\mu \la^a u + \bar{d} \ga^\mu \la^a d \rt]
    +\sqrt{\fr{8}{3}} \fr{g_1^2\de}{12} B'_\mu \!
     \lt[ \bar{u} \ga^\mu (5 + 3\ga_5) u
         -\bar{d} \ga^\mu (1 + 3\ga_5) d
     \rt] ,
\eql{hrho-q-qbar:Rhadron}
\eeq
where $g_3$ and $g_1$ are the SM $\SUC$ and $\UY$ couplings, respectively, and $\la^a$ ($a=1,\cdots,8$) are the Gell-Mann matrices. There are similar
couplings for the 2nd and 3rd generations with exactly the same coupling strength. The parameter $\de$ cannot be precisely determined and we use
\eq{delta} as an estimate in our computations below, bearing in mind that the results will have $\cO(1)$ uncertainties. Once a $\hrho$ is produced, its
decay is described by
\beq
  \cL_\text{int}
  &=& -\fr{ig_\hrho}{2} G'^{a}_\mu
       \lt[ \hpi_3^\dag \la^a (\del_\mu \hpi_3) - (\del_\mu \hpi_3^\dag) \la^a \hpi_3 \rt]
      +g_\hrho f^{abc}\, G'^{a}_\mu \, \hpi_8^b (\del_\mu \hpi_8^c)  \nn\\
   && +i\sqrt{\fr23} g_\hrho \, B'_\mu
       \lt[ \hpi_3^\dag (\del_\mu \hpi_3) - (\del_\mu \hpi_3^\dag) \hpi_3 \rt]     \,,
\eeq
where $g_\hrho$ here cannot be precisely determined and we use \eq{ghrho} as an estimate, which we expect to hold up to an $\cO(1)$ uncertainty.

\DOUBLEFIGURE[t]{plot_Rhadronproductionxsec.eps, width=1.0\linewidth} {plot_Rhadron_shat.eps, width=1.0\linewidth} {\figl{totalxsec_Rhadron}The
pair-production cross-section for the $\hpi_3$ at the LHC ($E_{\rm CM}=14~{\rm TeV}$). Vanishing hyperquark masses are assumed for the ratio of $m_{\hpi}
/ m_{\hrho}$.} {\figl{invmass_Rhadron}The invariant mass distribution of the $\hpi_3$ pairs for mass points 1 through 4.}

The $\hpi_3$ and $\hpi_8$ can also be pair-produced from a $q$-$\bar{q}$ initial state via an $s$-channel $g$
(or via $Z$ or $\ga$, which are negligible).
This is described by
\beq
  \cL_\text{int}
  &=& -\fr{ig_3}{2} G^{a}_\mu
       \lt[ \hpi_3^\dag \la^a (\del_\mu \hpi_3) - (\del_\mu \hpi_3^\dag) \la^a \hpi_3 \rt]
      -g_3 f^{abc}\, G^{a}_\mu \, \hpi_8^b (\del_\mu \hpi_8^c)  \nn\\
   && +\fr{4ig_1}{3} (A^\mu \cos\tht - Z^\mu \sin\tht) \lt[ \hpi_3^\dag (\del_\mu \hpi_3) - (\del_\mu \hpi_3^\dag) \hpi_3 \rt]  \,,
\eeq
where $\la^a$ ($a=1,\cdots,8$) are the Gell-Mann matrices and $f^{abc}$ is the $\SU(3)$ structure constant normalized as $[\la^a, \la^b] =2if^{abc}
\la^c$. Moreover, since the $\hpi_3$ and $\hpi_8$ are colored, they can also be produced from a $g$-$g$ initial state as in \Fig{gg-to-pipi}. To
accurately describe this process one needs to augment the above terms by the following quartic vertices
\beq
  \cL_\text{int}
  = \fr{g_3^2}{4} \, G^a_\mu G^{b\mu} \, \hpi_3^\dag \la^a \la^b \hpi_3
   +\fr{g_3^a}{2} f^{abc}f^{ade} \, G^{b}_\mu G^{d\mu} \, \hpi_8^c \hpi_8^e  \,.
\eeq
Similar 4-point vertices with the $\UY$ gauge boson $B_\mu$ are irrelevant for LHC phenomenology and we omit them here.

\DOUBLEFIGURE[t]{plot_Rhadron_eta.eps, width=1.0\linewidth} {plot_Rhadron_beta.eps, width=1.0\linewidth} {\figl{eta_Rhadron}The rapidity distribution for
R-hadron pair production at mass points 1 through 4.} {\figl{beta_Rhadron}The velocity distribution for R-hadron pair production at mass points 1 through
4.}

The $\hpi_3$ is stable at collider time scales, forming an R-hadron after QCD hadronization. (For the interested reader, we discuss the eventual decay of
$\hpi_3$ in \App{Rhadron}.) On the other hand, the $\hpi_8$ promptly decays to $gg$, $gZ$ and $g\ga$, which is described by
\beq
  \cL_\text{int}
  &=& -\fr{3 g_3^2}{64\pi^2 f_\hpi} \ep^{\mu\nu\rho\sg} \mathop{\rm tr}[\la^a \{ \la^b, \la^c\}] \,
       \hpi_8^a (\del_\mu G^b_\nu) (\del_\rho G^c_\sg)  \nn\\
   && -\fr{g_1 g_3 \sin\tht}{4\pi^2 f_\hpi} \ep^{\mu\nu\rho\sg} \,
       \hpi_8^a (\del_\mu G^a_\nu) (\del_\rho Z_\sg)
      +\fr{g_1 g_3 \cos\tht}{4\pi^2 f_\hpi} \ep^{\mu\nu\rho\sg} \,
       \hpi_8^a (\del_\mu G^a_\nu) (\del_\rho A_\sg)  \,,
\eeq
where $\tht$ is the weak mixing angle. The hyperpion decay constant $f_\hpi$ can be estimated using the analogy with QED-QCD as $f_\hpi \sim f_\pi m_\hrho
/ m_\rho$, which is $\cO(100)$ GeV. One sees that the $\hpi_8$ decay is dominated by the $gg$ final state, the $gZ$ and $g\ga$ branching fractions being
less than a percent. Therefore the $\hpi_8$ is practically identical to the color-octet hyperpion studied in Refs.\ \cite{Coloron-1, Coloron-2},
which always decays to $gg$. As shown in Ref.\ \cite{Coloron-2}, one can make a very strong case for reconstructing the $\hpi_8$ from
a multi-jet final state, although
reconstructing the $g'$ is harder, as the pair production of $\hpi_8$ is dominated by the $g$-$g$ initial state, rather than $q$-$\bar{q}$ where the $g'$
appears as a resonance. In this benchmark model, the existence of the R-hadrons will significantly facilitate discovery, so we will not repeat the
analysis in the multi-jet final state. In \Fig{totalxsec_Rhadron}, we show the total cross-section for the $\hpi_3$ pair production as a function of
$m_{\hpi_3}$, assuming vanishing hyperquark masses (i.e.\ the largest mass gap between $\hpi_3$ and $g'$) to be conservative.

As in our study of the other benchmark model, we begin by choosing a few mass points. These are
\beq
\begin{tabular}{|c||c|c|c|}
\hline
mass point & $m_{\hrho}$ (TeV)  & $m_{\hpi_3}$ (GeV)  & $m_{\hpi_8}$ (GeV)\\
\hline
1          & $1.5$              & $300$              & $435$ \\
2          & $1.5$              & $600$              & $800$ \\
3          & $1.0$              & $300$              & $435$ \\
4          & $2.5$              & $500$              & $725$ \\
\hline
\end{tabular}
\eeq
Mass points 1 and 4 are the most conservative as they maximize the $\hpi$-$\hrho$ mass gap (corresponding to zero hyperquark masses),
while mass points 3 and 2 are chosen to represent cases with light and heavy $\hpi$, respectively,
where the mass gap between the $\hpi$s and $\hrho$s is reduced due to nonzero hyperquark masses.
In \Fig{invmass_Rhadron} we show the invariant mass distribution of the $\hpi_3$ pair for these mass points.

\DOUBLEFIGURE[t]{plot_Rhadron_tofearly.eps, width=1.0\linewidth} {plot_Rhadron_toflate.eps, width=1.0\linewidth} {\figl{tofearly_Rhadron}The time lag for
the first R-hadron to arrive at the muon system at mass points 1 through 4.} {\figl{toflate_Rhadron}The time lag for the second R-hadron to arrive at the
muon system at mass points 1 through 4.}

Note that we have substantial production rates because we are producing colored states.
Furthermore, the mass gap between the $\hpi_3$ and $\hrho$ in this
benchmark is smaller than the gap between the $\hpiD$ (CHAMP) and $\hrho$ in the di-CHAMP benchmark, i.e., for the same $\hpi$ mass, the $\hrho$ in this
benchmark is lighter which enhances the resonant component to the production. Both these factors help make the reconstruction of the parent $\hrho$
resonance from R-hadron feasible at the LHC.  The plots in the section are made at parton-level, and that there may be ${\mathcal O}(1)$
inefficiencies due to hadronization and detector effects. As in our study of the CHAMP final state, such effects should be studied with a full detector
simulation, and we leave that task to a study by experimentalists.  But, thanks to the large rates, our crude analyses should be sufficient for
demonstrating the discovery potential in this channel.

We proceed as in our study of CHAMPs to study general kinematic features and prospects of triggering. In \Fig{eta_Rhadron} we plot the rapidity
distribution of the R-hadrons for mass points 1 through 4. Note that while the production is still concentrated in the central region, the peak of the
distribution is not exactly at zero, which is an indication that the gluon initial states have a significant contribution to the cross section. We find
that both R-hadrons will be within acceptance of the muon system ($|\eta|<2.5$) with $80\%$ efficiency at mass point 1, $95\%$ at mass point 2, $87\%$ at
mass point 3 and $86\%$ at mass point 4. In \Fig{beta_Rhadron} we plot the velocity distribution of the R-hadrons. Similar to the case of CHAMPs one sees
that the velocities are not peaked at threshold; note however that for mass points 2 and 3 where the resonance channel is not too far from kinematic
threshold, there is a fraction of smaller velocity events. The production being away from threshold is once again good news for triggering. Using the same
definitions as in our study of the CHAMP final state, we plot in \Fig{tofearly_Rhadron} and \Fig{toflate_Rhadron} the time lag for the first and second
R-hadron to reach the muon system, where triggering occurs. The majority of events have time lags less than $25$ ns which should yield a high trigger
efficiency.

\EPSFIGURE[t]{plot_4Rhadron_xsec.eps, width=0.8\linewidth} {\figl{4Rhadrons}The production cross-section of four R-hadrons through the decay of two
intermediate $g'$s, as a function of the $g'$ mass at the LHC ($E_{\rm CM}=14$ TeV). We use three values for the parameter $\al$, which is expected to be
close to 1. Vanishing hyperquark masses are assumed.}

While there are various other models of physics beyond the SM that have collider stable colored particles, we wish to point out a feature of the R-hadrons
in vectorlike confinement that is (to the best of our knowledge) unique: the production of {\em four} $\hpi_3$s from the decays of a {\em pair} of $g'$s!

Parameterizing the $g'$ pair-production is a tricky issue.  For the resonant $g'$ production, since it is a combination of resonant production and
two-body decay, it is sufficient to parameterize the 2-to-1 production rate, the Breit-Wigner propagator for the intermediate resonance, and the two-body
decay rate; in the narrow-width approximation, this is indeed a complete parametrization. On the other hand, for the pair production of $g'$s, the angular
distribution of the pair must be parameterized in addition to the production rate. However, the momentum transfer involved in $q\bar{q}$, $gg \to g'g'$ is
$\sim\cO(m_\hrho)$, which is the inverse of the size of the bound state itself. Thus, there is no good field theoretic way to parameterize the angular
distribution, nor can we analytically determine the full form factor for the $g'$ pair production due to strong coupling. Therefore, instead of an
analysis of detailed kinematic features of this final state, we will give an estimate of the overall rate of 4 R-hadron production, and show that it is
large enough for these events to be observable even when one allows for ${\mathcal O}(1)$ uncertainties and efficiencies due to acceptance.

With this cautionary remark in mind, we include the following terms in the phenomenological lagrangian to parameterize the $g'$ pair production:
\beq
  \cL_\text{int}
  &=&  g_3 f^{abc} \, (\del_\mu G'^a_\nu) (G^{b\mu} G'^{c\nu} - G^{b\nu} G'^{c\mu})
      -\fr{g_3^2}{4} f^{abc} f^{ade} \, G^b_\mu G'^c_\nu (G^{d\mu} G'^{e\nu} - G^{d\nu} G'^{e\mu})  \nn\\
   && +\al g_3 f^{abc} \, (\del_\mu G^a_\nu) G'^{b\mu} G'^{c\nu}
      -\fr{\al g_3^2}{2} f^{abc} f^{ade} G^{b}_\mu G^{c}_\nu G'^{d\mu} G'^{e\nu}  \,,
\eeq
where the terms in the first line come from the gauge-invariant kinetic term for $g'$, $\mathop{\rm tr}[G'_{\mu\nu} G'^{\mu\nu}]$, where $G'_{\mu\nu} =
[D_\mu, G'_\nu] - [D_\nu, G'_\mu]$ with $D_\mu = \del_\mu + ig_3 G^a_\mu \la^a/2$.  The terms in the second line come from $\mathop{\rm tr}[G^{\mu\nu}
[G'_\mu, G'_\nu]]$, which is gauge-invariant by itself and can thus have an independent coefficient, $\al$.
We use $\al$ to parameterize our $\cO(1)$ ignorance of the $g'$ pair-production rate.
Since the hypercolor sector contains no small or large parameters, we expect that $\al \sim \cO(1)$.
For more discussions on the above effective vertices, see Refs.\ \cite{Coloron-1, Coloron-2}.

Using the above effective vertices as well as the $g'\rightarrow \hpi_{3}\hpi_{3}$ branching fraction, we plot in \Fig{4Rhadrons} the production
cross-section of four R-hadrons via two intermediate $g'$s for representative values of $\al$. Negligible hyperquark masses are assumed in order to be
maximally conservative about the R-hadron-$g'$ mass gap.  We only plot $g'$ masses above a TeV, as smaller masses lead to R-hadrons that are too light for
Tevatron exclusion bounds. Note that even for fairly large values of the $g'$ mass, the rate of 4 R-hadron events is non-negligible. Along with R-hadron
pair production and multi-jet resonance signals, this novel final state should provide a strong indication for vectorlike confinement as the underlying
dynamics.

%%%%%%%%%%%%%%%%%
\section{Summary}
\Secl{conc}
%%%%%%%%%%%%%%%%%
We have studied the collider signatures of the most representative final states of the scenario of vectorlike confinement, where new fermions charged
under the SM form various bound states, hyperhadrons, due to a new confining gauge interaction, hypercolor.
Among them, those that are particularly important for LHC phenomenology are spin-1 hypermesons
that can mix with the SM gauge bosons,
pseudoscalar hypermesons that decay promptly to a pair of SM gauge bosons as well as those that are stable on
collider time scales.
We have focused on processes where a pair of pseudoscalars are produced either through ``Drell-Yan''
or through the resonant production and subsequent decay of one of the spin-1 hypermesons.

After giving a brief overview of the framework of vectorlike confinement and generic states that appear in it, we proceeded to look at specific cases
which represent the most common signatures of vectorlike confinement. In particular, we have focused on two benchmark models for two qualitatively
different sets of signatures; the new particles in one benchmark are charged under SM color, while those of the other benchmark are charged under the
electroweak interactions only. For the latter, the distinctive final states are pairs of stable charged particles (CHAMPs) and events with three photons.
For the colored benchmark we highlighted the presence of stable heavy hadrons (R-hadrons) in addition to the resonances in multi-jet channels studied in
Refs.\ \cite{Coloron-1, Coloron-2}.

For the CHAMP benchmark, we have given the details of the spectrum, which contains an electroweak-doublet pseudoscalar, a component of which
is a CHAMP, as well as a triplet pseudoscalar which decays to
pairs of photons, $W$s and $Z$s. These states are light compared to the mass of the spin-1 hypermeson, which is
around the confinement scale of hypercolor $\sim \cO(1)$ TeV.
The spin-1 hypermesons in this model can mix with $\ga$, $W^{\pm}$ and $Z$. We have written down an effective
Lagrangian that can be used to parameterize the relevant production and decays. Choosing three mass points, we have first studied the kinematic features
of the CHAMP final state, in particular showing that the triggering efficiency is expected to be high.
Since irreducible SM backgrounds are not present,
we have shown that kinematic reconstruction can point to the existence of the spin-1 hypermeson in the invariant mass distribution of the CHAMPs,
distinguishing vectorlike confinement from other theories with CHAMPs.
We then studied the $3\ga+W$ final state, showing that the triplet scalar can be
reconstructed easily from two of the photons, and the other side of the event can be reconstructed from the knowledge of the diphoton resonance mass.

For the colored benchmark, we have once again dwelled on the spectrum and written an effective Lagrangian to parameterize the relevant processes. We then
focused our attention on the R-hadron final states, as the multi-jet resonances have been studied in Refs.\ \cite{Coloron-1,Coloron-2}. As in the case of
CHAMPs, we have studied the kinematics of this final state for four different mass points and we have shown that a significant fraction of the R-hadrons
are produced with high enough velocity such that triggering is not a problem, and the large production cross-sections due to QCD makes this an easy final
state to look for. Again, the invariant mass distributions of R-hadron pairs exhibit a feature due to the spin-1 resonance, reflecting the underlying
vectorlike confinement.  Furthermore, we have highlighted a signal that can be uniquely attributed to vectorlike confinement, namely the production of
four R-hadrons via two intermediate spin-1 hypermesons. We have shown that the rate for this novel process is sizable at the LHC, providing another probe
of vectorlike confinement.

While we have not attempted to give rigorous background estimates for processes where detector effects become the main contribution to background, we have
argued that our analyses
should be sufficient to demonstrate the discovery potential of these final states at the LHC. It is our hope that a more
thorough study will be performed by experimentalists using full detector simulation, to improve upon our results and determine the precise
reach of the LHC in the parameter space of our benchmark models.

%%%%%%%%%%%%%%%%
\acknowledgments
%%%%%%%%%%%%%%%%
The authors would like to thank for Neil Christensen for his infinite patience in helping us use CalcHep, Johan Alwall for helping us with the Pythia-PGS
interface of MadEvent, Yuri Gershtein for his guidance and advice in the analysis of the $3\ga$ final state, Todd Adams and Jie Chen for their input on
the CHAMP analysis, and Raman Sundrum for fruitful discussions in the early stage of this work. We also thank T.\ Adams and R.\ Sundrum for comments on
the manuscript. C.K.\ is supported by DOE grant DE-FG02-96ER50959. T.O.\ is supported by his start-up funds at Florida State University.

%%%%%%%%%
%%%%%%%%%
\appendix
%%%%%%%%%
%%%%%%%%%

%%%%%%%%%%%%%%%%%%%%%%%%%%%%%%%%%%%%%%%%%%%%%%%%%%%%%%%%%%%%%
\section{The Fundamental Lagrangians of the Benchmark Models}
\Appl{fundamental}
%%%%%%%%%%%%%%%%%%%%%%%%%%%%%%%%%%%%%%%%%%%%%%%%%%%%%%%%%%%%%

%%%%%%%%%%%%%%%%%%%%%%%%%%%%%%%%%%%%%%%%%%%%%%%
\subsection{The Di-CHAMP/Multiphoton Benchmark}
\Appl{dichamp}
%%%%%%%%%%%%%%%%%%%%%%%%%%%%%%%%%%%%%%%%%%%%%%%
The fundamental Lagrangian defined at a scale far above the hypercolor confinement scale ($\sim \cO(1)$ TeV) reads
\beq
\hspace{-1em}
 \cL = \cL_\text{SM}
      -\fr14 H^a_{\mu\nu} H^{a\mu\nu}
      +\wba{\psi}_1 i\Sla{D} \psi_1 - m_1 \wba{\psi}_1 \psi_1
      +\wba{\psi}_2 i\Sla{D} \psi_2 - m_2 \wba{\psi}_2 \psi_2
      +\fr{\tht_H}{4} \ep^{\mu\nu\rho\sg} H^a_{\mu\nu} H^a_{\rho\sg} \,.
\eeq
Here, $H^a_{\mu\nu}$ ($a=1$, $\cdots$, 8) is the field strength of the hyper-color gauge field $H^a_\mu$ with the hypercolor group chosen to be $\SU(3)$.
$\psi_1$ and $\psi_2$ are Dirac fermions transforming as follows:
\beq
\begin{tabular}{|c||c|c|c|c|}
\hline
         & $\SU(3)_\text{HC}$ &    $\SUC$ &    $\SUL$ & $\UY$ \\
\hline
$\psi_1$ &          ${\bf 3}$ & ${\bf 1}$ & ${\bf 1}$ &  $-1$ \\
$\psi_2$ &          ${\bf 3}$ & ${\bf 1}$ & ${\bf 2}$ & $1/2$ \\
\hline
\end{tabular}
\eeq
The above Lagrangian is the most general renormalizable Lagrangian for this field content. Note that if we turn off SM gauge interactions, this is a
theory with three hypercolors and three ``hyperflavors'', exactly analogous to the low energy QCD, which is a theory with three colors and three flavors.
Therefore, this benchmark model has the theoretical advantage of being calculable by using QCD as an ``analog computer''.

The axial currents that can create $\hpi$s are as follows.  The $\hpiT$ can be created by
\beq
  J_{\rm 5T}^{\mu a} = \wba{\psi}_2 \ga^\mu \ga_5 \sg^a \psi_2 \,,
\eeq
where $a=1$, 2, 3, and $\sg^a$ are the Pauli matrices acting on the $\SUL$ doublet $\psi_2$. On the other hand, the $\hpiD$ can be created by
\beq
  J_{\rm 5D}^\mu = \wba{\psi}_1 \ga^\mu \ga_5 \psi_2 \,.
\eeq
From this, one sees that a $\hpiD$ carries a nonzero $\psi_1$ number (and $\psi_2$ number), leading to its long lifetime. Finally, the $\hpiS$ can be
created by
\beq
  J_{\rm 5S}^\mu = \wba{\psi}_2 \ga^\mu \ga_5 \psi_2 - 2 \wba{\psi}_1 \ga^\mu \ga_5 \psi_1  \,.
\eeq

In addition to the above renormalizable interactions, the Lagrangian should also contain nonrenormalizable interactions responsible for the decays of the
long-lived hyper-pions (i.e.\ $\hpiD^\pm$) and hyper-baryons on the cosmological time scale. The $\hpiD$ can decay via a nonrenormalizable operator
\beq
  \cL_\text{$\hpiD$ decay}
  = \fr{c_{ij}}{M^2} J_{\rm 5D}^\mu \, e_{Ri}^{\rm T} \cC \ga_\mu \ell_{Lj}  \,,
\eeq
where $i,j=1, 2, 3$ are generation indices, and $\cC$ is a charge conjugation matrix satisfying $\cC \ga_\mu \cC^{-1} = -\ga_\mu^{\rm T}$.  As discussed
in Ref.\ \cite{VC}, the absence of excessive flavor/CP violations beyond the SM implies $M \gsim 10^4$ TeV for generic $\cO(1)$ $c_{ij}$, rendering the
$\hpiD$ stable on collider time scales.

For hyperbaryon decays, let us focus on how spin-$1/2$ hyperbaryons decay, as higher spin hyperbaryons are heavier and will quickly decay to a spin-$1/2$
hyperbaryon.  Spin-$1/2$ hyperbaryons can be created by the operators
\beq
   B_{1L} = (\psi_{2L}^{\rm T} \cC \psi_{2L}) \psi_{2L}  &\sim& ({\bf 1}, {\bf 2}, 3/2)  \,,\nn\\
   B_{2L,1} = (\psi_{2L}^{\rm T} \cC \psi_{2L}) \psi_{1L}  &\sim& ({\bf 1}, {\bf 1}, 0)  \,,\nn\\
   B_{2L,2} = (\psi_{2L}^{\rm T} \cC \psi_{1L}) \psi_{2L}  &\sim& ({\bf 1}, {\bf 1}, 0)  \,,\nn\\
   B_{3L} = (\psi_{2L}^{\rm T} \cC \psi_{1L}) \psi_{2L}  &\sim& ({\bf 1}, {\bf 3}, 0)  \,,\nn\\
   B_{4L} = (\psi_{2L}^{\rm T} \cC \psi_{1L}) \psi_{1L}  &\sim& ({\bf 1}, {\bf 2}, -3/2)  \,,
\eeq
and their right-handed counterparts. If stable, they would lead to cosmological problems. However, just like the proton in the SM, their stability is
susceptible to nonrenormalizable operators that violate hyperbaryon number. In fact, they can all decay to SM particles via
\beq
   B_{1R}^{\rm T} \cC e_{Ri} H^*  \,,\qquad
   \wba{B}_{2R,i} \ell_{Lj} H  \,,\qquad
   \wba{B}_{3R} \ell_{Li} H  \,,\qquad
   \wba{B}_{4L} e_{Ri} H^* \,,
\eeq
where $H$ is the SM Higgs field. Note that the mass scale suppressing these operators has no reason to be the same as the scale suppressing the $\hpiD$
decay operator, since the latter operator preserves hyperbaryon number while the former operators do not.

%%%%%%%%%%%%%%%%%%%%%%%%%%%%%%%%%%%%%%
\subsection{The Di-R-Hadron Benchmark}
\Appl{Rhadron}
%%%%%%%%%%%%%%%%%%%%%%%%%%%%%%%%%%%%%%
The fundamental Lagrangian defined at a scale far above the hypercolor confinement scale ($\sim \cO(1)$ TeV) reads
\beq
\hspace{-1em}
 \cL = \cL_\text{SM}
      -\fr14 H^a_{\mu\nu} H^{a\mu\nu}
      +\wba{\psi}_1 i\Sla{D} \psi_1 - m_1 \wba{\psi}_1 \psi_1
      +\wba{\psi}_3 i\Sla{D} \psi_3 - m_3 \wba{\psi}_3 \psi_3
      +\fr{\tht_H}{4} \ep^{\mu\nu\rho\sg} H^a_{\mu\nu} H^a_{\rho\sg} \,.
\eeq
Here, $H^a_{\mu\nu}$ ($a=1$, $\cdots$, 8) is the field strength of the hyper-color gauge field $H^a_\mu$ with the hypercolor group chosen to be $\SU(3)$.
$\psi_1$ and $\psi_3$ are Dirac fermions transforming as follows:
\beq
\begin{tabular}{|c||c|c|c|c|}
\hline
         & $\SU(3)_\text{HC}$ &    $\SUC$ &    $\SUL$ &  $\UY$ \\
\hline
$\psi_1$ &          ${\bf 3}$ & ${\bf 1}$ & ${\bf 1}$ &    $1$ \\
$\psi_3$ &          ${\bf 3}$ & ${\bf 3}$ & ${\bf 1}$ & $-1/3$ \\
\hline
\end{tabular}
\eeq
The above Lagrangian is the most general renormalizable Lagrangian for this field content. Note that if we turn off SM gauge interactions, this is a
theory with three hypercolors and {\em four} ``hyperflavors''. So, unlike the Di-CHAMP benchmark model, this is not exactly analogous to the low-energy
QCD.  But we still expect that many numbers can be extracted from the low-energy QCD analogy with only $\cO(1)$ uncertainties.

The axial currents that can create $\hpi$s are as follows.  The $\hpi_8$ can be created by
\beq
  J_{\rm 5,8}^{\mu a} = \wba{\psi}_3 \ga^\mu \ga_5 \la^a \psi_3 \,,
\eeq
where $a=1$, $\cdots$, 8, and $\la^a$ are the Gell-Mann matrices acting on the color-triplet $\psi_3$. On the other hand, the $\hpi_3$ can be created by
\beq
  J_{\rm 5,3}^\mu = \wba{\psi}_1 \ga^\mu \ga_5 \psi_3 \,.
\eeq
From this, one sees that a $\hpi_3$ carries a nonzero $\psi_1$ number (and $\psi_3$ number), leading to its long lifetime. Finally, the $\hpi_1$ can be
created by
\beq
  J_{\rm 5,1}^\mu = \wba{\psi}_3 \ga^\mu \ga_5 \psi_3 - 3 \wba{\psi}_1 \ga^\mu \ga_5 \psi_1  \,.
\eeq

In addition to the above renormalizable interactions, the Lagrangian should also contain nonrenormalizable interactions responsible for the decays of the
long-lived hyper-pions (i.e.\ $\hpi_3$) and hyper-baryons on the cosmological time scale. For the $\hpi_3$ decay, note that the pseudoscalar operator,
\beq
  P_3 = \wba{\psi}_1 \ga_5 \psi_3  \,,
\eeq
can also create a $\hpi_3$. Using this, we can write down a nonrenormalizable operator that leads to $\hpi_3$-decay:
\beq
  \cL_\text{$\hpi_3$ decay}
  = \fr{c_{ij}}{M^2} P_3 \, d_{Ri}^{\rm T} \cC e_{Rj}  \,,
\eeq
where $i,j=1$, 2, 3 are generation indices, and $\cC$ is the charge conjugation matrix satisfying $\cC \ga_\mu \cC^{-1} = -\ga_\mu^{\rm T}$. As discussed
in Ref.\ \cite{VC}, the absence of excessive flavor/CP violations beyond the SM implies $M \gsim 10^4$ TeV for generic $\cO(1)$ $c_{ij}$, rendering the
$\hpi_3$ stable on the collider time scale.

For hyperbaryon decays, let us focus on how spin-$1/2$ hyperbaryons decay, as higher spin hyperbaryons are heavier and will quickly decay to a spin-$1/2$
hyperbaryon.  Spin-$1/2$ hyperbaryons can be created by the operators
\beq
   B_{1L}   = (\psi_{3L}^{\rm T} \cC \psi_{3L}) \psi_{3L}  &\sim& ({\bf 8},       {\bf 1}, -1)  \,,\nn\\
   B_{2L}   = (\psi_{3L}^{\rm T} \cC \psi_{3L}) \psi_{3L}  &\sim& ({\bf 1},       {\bf 1}, -1)  \,,\nn\\
   B_{3L,1} = (\psi_{3L}^{\rm T} \cC \psi_{3L}) \psi_{1L}  &\sim& ({\bf \wba{3}}, {\bf 1}, 1/3)  \,,\nn\\
   B_{3L,2} = (\psi_{3L}^{\rm T} \cC \psi_{1L}) \psi_{3L}  &\sim& ({\bf \wba{3}}, {\bf 1}, 1/3)  \,,\nn\\
   B_{4L}   = (\psi_{3L}^{\rm T} \cC \psi_{1L}) \psi_{3L}  &\sim& ({\bf 6},       {\bf 1}, 1/3)  \,,\nn\\
   B_{5L}   = (\psi_{3L}^{\rm T} \cC \psi_{1L}) \psi_{1L}  &\sim& ({\bf 3},       {\bf 1}, 5/3)  \,,
\eeq
and their right-handed counterparts. If stable, they would lead to cosmological problems. However, just like the proton in the SM, their stability is
susceptible to nonrenormalizable operators that violate hyperbaryon number. In fact, they can all decay to SM particles via
\beq
  && \wba{e}_{Ri}[\ga^\mu,\ga^\nu] B_{1L}^a \, G_{\mu\nu}^a  \,,\qquad
     \wba{e}_{Ri} B_{2L}  \,,\qquad
     B_{3R,i}^{\rm T} \cC q_{Lj} H^*  \,,\nn\\
  && (\wba{B}_{4L} u_{Ri}) \, (u_{Rj}^{\rm T} \cC e_{Rk})   \,,\qquad
     (\wba{u}_{Ri} B_{5L}) (\ell_{Lj}^{\rm T} \cC \ell_{Lk})  \,,
\eeq
where $G_{\mu\nu}^a$ ($a=1$, $\cdots$, 8) is the gluon field strength. Note that the mass scale suppressing these operators has no reason to be the same
as the scale suppressing the $\hpi_3$ decay operator, since the latter operator preserves hyperbaryon number while the former operators do not.

%%%%%%%%%%%%%%%%%%%%%%%%%%%
%%%%%%%%%%%%%%%%%%%%%%%%%%%

\end{document}